\documentclass[aps,pra,reprint,groupedaddress,amsmath,amssymb]{revtex4-2}

\usepackage{amsmath}
\usepackage{amssymb}
\usepackage{amsfonts}
\usepackage{amsthm}
\usepackage{amscd}
\usepackage{mathrsfs}
\usepackage{bm}

\usepackage{graphicx}
\usepackage{rotating}

\usepackage{array}
\usepackage{multirow}
\usepackage{booktabs}

\newtheoremstyle{paperstatement}{6pt}{6pt}{\normalfont}{}{\bfseries}{.}{.5em}{}
\theoremstyle{paperstatement}
\newtheorem{theorem}{Theorem}
\newtheorem{proposition}{Proposition}
\newtheorem{corollary}{Corollary}

\usepackage{hyperref}
\makeatletter
\def\@email#1#2{\endgroup\@AF@join{#1#2}}
\makeatother
\hypersetup{
  colorlinks=true,
  linkcolor=blue,
  citecolor=blue,
  urlcolor=blue
}

\newcommand{\ii}{\mathrm{i}}
\newcommand{\dd}{\mathrm{d}}
\newcommand{\sech}{\operatorname{sech}}
\newcommand{\sgn}{\operatorname{sgn}}

\begin{document}

\title{Realizability and tunability of synchronized optical soliton collision patterns}
\author{Bo-Jie Deng}
\author{Rui Guo}
\email[Contact author: ]{gr81@sina.com}
\author{Jia-Xue Niu}
\author{Jian-Wen Zhang}
\affiliation{School of Mathematics, Taiyuan University of Technology, Taiyuan, Shanxi 030024, China}

\date{\today}

\begin{abstract} Synchronizing the positions and phases of colliding optical solitons can maximize the amplitude at the collision center, producing a large-amplitude localized wave. We investigate whether multiple prescribed synchronized collisions can be jointly realized within a single multisoliton field and to what extent their relative retarded times and propagation distances can be varied continuously while preserving synchronization. Ideal optical fiber propagation is modeled here by the focusing nonlinear Schr\"odinger equation with zero boundary conditions. For distinct discrete eigenvalues, we derive an affine system of equations constraining the collision event positions to determine whether the synchronization conditions at different events can be satisfied simultaneously. Furthermore, for a fixed discrete spectrum and a realizable complete collision organization in which each soliton pair participates in exactly one event, continuous tuning of the relative event positions while preserving synchronization requires collinearity of the auxiliary spectral points constructed from the discrete eigenvalues. When these points are noncollinear, only uniform translations of all event centers are possible. Any three-soliton spectrum with distinct soliton velocities and noncollinear auxiliary spectral points supports both a collective collision and a rigid fully pairwise collision organization. A physically admissible four-soliton family with collinear auxiliary spectral points demonstrates continuous internal tunability.
\end{abstract}

\maketitle

\section{Introduction}
\label{sec:introduction}

Extreme optical events characterized by strong localization and pronounced transient amplitude enhancement have attracted considerable interest in nonlinear optics~\cite{MussotEtAl2009,Kibler2010,Dudley2014}. Experimental and theoretical studies have revealed various such events and motivated investigations of the nonlinear dynamics underlying their formation~\cite{Solli2007,DudleyEtAl2019}.  The focusing nonlinear Schr\"odinger (NLS) equation provides a fundamental envelope model for investigating such phenomena in optical fibers, where anomalous group velocity dispersion and Kerr nonlinearity govern pulse evolution~\cite{HasegawaTappert1973}. In dimensionless variables, it reads \begin{equation}
  \ii q_t+q_{xx}+2|q|^2q=0,
  \label{eq:nls-general}
\end{equation}
where $q(x,t)$ is the normalized complex optical envelope, $x$ denotes retarded time in a comoving reference frame, and $t$ denotes the normalized propagation distance.

Within this model, a widely studied mechanism for generating such large-amplitude localized waves is modulation instability, which amplifies small perturbations of a continuous-wave background and can lead to rogue-wave formation during the subsequent nonlinear evolution~\cite{BenjaminFeir1967,Tai1986,OnoratoEtAl2001,ZakharovOstrovsky2009,Onorato2013,Dudley2014,DudleyEtAl2019}. The Peregrine soliton, a doubly localized rational solution, provides a prototypical model of rogue waves~\cite{Peregrine1983,AkhmedievEtAl1987,DystheTrulsen1999, ShriraGeogjaev2010,Kibler2010,Chabchoub2011}. Beyond this fundamental solution, higher-order rational generalizations describe the complex dynamics of multiple interacting rogue waves~\cite{AkhmedievAnkiewiczTaki2009,DubardEtAl2010,GuoLingLiu2012}. In the maximally compressed state, an $n$th-order solution can attain a peak amplitude of $2n+1$ times the background amplitude~\cite{Akhmediev2009,OhtaYang2012,WangEtAl2017}. Interestingly, these higher-order structures possess internal free parameters whose variation can reorganize the central compressed structure into different spatiotemporal patterns of peaks locally approximated by fundamental Peregrine solitons, possibly accompanied by a lower-order structure near the pattern center~\cite{OhtaYang2012,DubardMatveev2013,Kedziora2011,Kedziora2012,Kedziora2013,YangYang2021}.

Extreme optical events, however, need not arise from modulation instability. Instead, a complementary mechanism for their formation emerges from coherent multisoliton interactions. Experiments have observed rogue-wave generation through nonlinear soliton collisions in mode-locked fiber lasers \cite{Peng2016}. Within the integrable focusing NLS equation with zero boundary conditions, suitably arranged multisoliton pulses can undergo strong self-compression \cite{AkhmedievEtAl1989,AkhmedievMitskevich1991}, while coherent focusing of bound states and soliton collisions can produce transient large-amplitude peaks \cite{Gelash2018,ZhangEtAl2026}. Remarkably, the local structure of such soliton collisions can closely approximate that of rational rogue waves. In particular, a specially designed collision of three equal-amplitude solitons with symmetric velocities and appropriate relative phases has been shown to closely approximate a suitably scaled second-order Peregrine soliton near the collision, including its spatial profile and the temporal evolution of its maximal amplitude~\cite{Gelash2018}. Related studies have also identified extreme events with approximately rational profiles in multisoliton bound-state models~\cite{AgafontsevGelash2021}. More generally, spatially localized multisoliton solutions can be constructed to approximate Peregrine rogue-wave dynamics over extended spacetime regions at large soliton number~\cite{AgafontsevEtAl2024}.

While these studies demonstrate that multisoliton interactions can produce strongly localized waves, achieving a fully synchronized collision, in which all participating solitons focus coherently at a prescribed spacetime point, requires specific conditions on their positions and phases. By introducing the asymptotic middle states of the solitons, Sun~\cite{Sun2016} formulated these synchronization conditions as linear relations among the position and phase parameters and proved that the resulting peak amplitude equals the sum of the individual soliton amplitudes. For large numbers of equal-amplitude solitons with well-separated velocities, Girotti et al.~\cite{Girotti2025} derived local sinc asymptotics near the synchronized collision and established the universality of this profile under suitable random-amplitude assumptions. Related aspects of soliton synchronization have also been explored in other integrable models~\cite{SlunyaevPelinovsky2016,Slunyaev2019,SlunyaevTarasova2022,TarasovaSlunyaev2023}.

However, coherent focusing at a single collision event represents only one aspect of multisoliton interactions, which can also exhibit a rich variety of spacetime dynamics~\cite{Gordon1983,MitschkeMollenauer1987,ZhaoEtAl2016}. In a parallel line of research, studies have shown that, on a modulationally unstable continuous-wave background, the positions and phases of breathers can be adjusted to control their interactions, including multiple collisions~\cite{AkhmedievSotoCrespoAnkiewicz2009, FrisquetEtAl2013,GelashXuKibler2022}. For bright solitons on a zero background, the corresponding question is how multiple collision events can be arranged at distinct spacetime points within a single multisoliton field. Understanding how to manipulate optical solitons through suitable parameter choices is therefore important for arranging such events in optical fibers.

Recent optical-fiber experiments have demonstrated control over the parameters and collision configurations of bright solitons. In these studies, inverse scattering theory has guided the selective control of individual soliton velocities~\cite{MucciEtAl2025} and provided a framework for interpreting three-soliton interactions that change from a localized collision region to three distinct pairwise collisions as the initial pulse positions are varied~\cite{CopieEtAl2023}. In this framework, the discrete eigenvalues determine the amplitudes and velocities of the constituent solitons, while the associated norming data encode their position and phase information. For well-separated solitons, this position and phase information can be interpreted in terms of individual pulse centers and carrier phases. For a given discrete spectrum, varying the norming data allows us to investigate different collision configurations without changing the intrinsic properties of the constituent solitons. These considerations motivate the study of whether suitable choices of soliton parameters can produce prescribed arrangements of synchronized collisions at several spacetime points within a single multisoliton field, analogous to the parameter-dependent spatiotemporal patterns of higher-order rogue waves on a nonzero background.

The simultaneous realization of these events imposes compatibility requirements beyond those of an individual collision. At fixed constituent amplitudes and frequency parameters, the position and phase data chosen to synchronize one event also enter the synchronization conditions for other events involving the same solitons. Thus, even if each prescribed collision can be realized separately, a common set of norming data may not realize them all within a single multisoliton solution. Even when the events can be realized together, their relative positions need not admit continuous variation while preserving synchronization. A further question, therefore, is to what extent the relative spacetime geometry of the event centers can vary continuously at a fixed discrete spectrum while preserving synchronization.

To address these questions, we consider the reflectionless $N$-soliton solutions of the focusing NLS equation with zero boundary conditions, constructed through the inverse scattering transform~\cite{ZakharovShabat1972,SatsumaYajima1974,AblowitzPrinariTrubatch2004}. Based on their determinant representation, we analyze the compatibility of multiple synchronization conditions and examine how the discrete spectrum and collision organization constrain joint realizability and the tunability of the relative event positions. The paper is organized as follows. Section~\ref{sec:multisoliton-design} introduces the multisoliton parametrization and develops the local synchronization and compatibility conditions for prescribed collision events. Section~\ref{sec:realizability-rigidity} establishes the realizability and rigidity results, presents three- and four-soliton examples, and examines how synchronization constraints on cluster separation and relative phase are reflected in the Fourier interference of the mixed flexible family. The appendices provide the supporting proofs, the construction and admissibility checks for the mixed four-soliton family, and the derivations and numerical details of the Fourier analysis.

\section{Multisoliton parametrization and collision design}
\label{sec:multisoliton-design}

\subsection{Multisoliton solution and soliton parametrization}

Under the zero boundary condition
\begin{equation}
  q(x,t)\rightarrow0,
  \qquad |x|\rightarrow\infty,
  \label{eq:zero-boundary-condition}
\end{equation}
the focusing NLS equation is solved by the Zakharov--Shabat inverse scattering transform (IST)~\cite{ZakharovShabat1972}.  Assume that the discrete spectrum consists of $N$ simple eigenvalues
\begin{equation}
  \kappa_j=\alpha_j+\ii\beta_j,
  \qquad \beta_j>0,
  \qquad j=1,\ldots,N,
  \label{eq:eigenvalues-general}
\end{equation}
The eigenvalues are assumed to be distinct, whereas their real parts, and hence the associated soliton velocities, need not be distinct. Define the $N\times N$ matrix $\bm{G}$ and the $(N+1)\times(N+1)$ matrix $\bm H$ by
\begin{gather}
  G_{jm}
  =\frac{
  e^{-(g_j^*+g_m)}+e^{g_j^*+g_m}
  }{\kappa_j^*-\kappa_m},
  \qquad 1\leq j,m\leq N,
  \label{eq:G-matrix}\\
   \bm{H}
  =
  \begin{pmatrix}
  0 & e^{g_1} & \cdots & e^{g_N}\\
  e^{-g_1^*} & & & \\
  \vdots & & \bm{G} & \\
  e^{-g_N^*} & & &
  \end{pmatrix},
  \label{eq:H-matrix}
\end{gather}
where the superscript $*$ denotes complex conjugation, and
\begin{equation}
  g_j(x,t)
  =-\ii\kappa_jx-2\ii\kappa_j^2t+\frac{c_j}{2},
  \label{eq:half-exponent}
\end{equation}
with the complex norming parameters
\begin{equation}
  c_j=s_j+\ii\vartheta_j.
  \label{eq:norming-exponents}
\end{equation}

In this setting, the $N$-soliton solution of Eq.~\eqref{eq:nls-general} admits the determinant representation~\cite{Sun2016}:
\begin{equation}
  q_N(x,t)=-2\ii\frac{\det\bm{H}}{\det\bm{G}}.
  \label{eq:N-soliton-determinant}
\end{equation}

The one-soliton solution associated with the eigenvalue $\kappa_j$ is obtained from Eq.~\eqref{eq:N-soliton-determinant} as
\begin{equation}
  q_j(x,t)
  =\mathcal{A}_j\sech\mathcal{X}_j(x,t)
  e^{\ii\mathcal{P}_j(x,t)},
  \label{eq:single-soliton}
\end{equation}
where
\begin{equation}
  \mathcal{A}_j=2\beta_j,
  \qquad
  V_j=-4\alpha_j
  \label{eq:single-amplitude-velocity}
\end{equation}
are the peak amplitude and velocity.  The real envelope coordinate and carrier phase are
\begin{subequations}\label{envelope-phase-coordinate}
  \begin{align}
  \mathcal{X}_j(x,t)
  &=2\beta_j(x-V_jt)+s_j,
  \label{eq:single-envelope-coordinate}\\
  \mathcal{P}_j(x,t)
  &=-2\alpha_jx-4(\alpha_j^2-\beta_j^2)t+\vartheta_j.
  \label{eq:single-carrier-phase}
\end{align}
\end{subequations}
The soliton trajectory is therefore
\begin{equation}
  x_j^{(0)}(t)=V_jt+d_j,
  \qquad
  d_j=-\frac{s_j}{2\beta_j}.
  \label{eq:bare-center}
\end{equation}

\subsection{Asymptotic position and phase shifts}

For pairwise distinct velocities, the $N$-soliton solution asymptotically decomposes into $N$ separated fundamental solitons as $t\to\pm\infty$. Each soliton preserves its amplitude $\mathcal A_j$ and velocity $V_j$, while acquiring finite position and carrier phase shifts through interactions with the other solitons~\cite{ZakharovShabat1972,SatsumaYajima1974,AblowitzPrinariTrubatch2004}.

Equal-velocity solitons, while still associated with distinct simple eigenvalues, are treated as a comoving subsystem in the local formulation below. To represent the pairwise contribution of soliton $m$ to these shifts, we define
\begin{equation}
\Omega_{jm}
=
\log\!\left(
\frac{\kappa_j-\kappa_m}
     {\kappa_j-\kappa_m^*}
\right),
\qquad j\ne m .
\label{eq:scattering-log-principal}
\end{equation}
We use the principal logarithm, with argument in $(-\pi,\pi]$, throughout. A consistent change of logarithm branches, $\Omega_{jm}\mapsto\Omega_{jm}+2\pi\ii k_{jm}$, $k_{jm}\in\mathbb Z$, leaves all real scattering shifts unchanged and does not affect the synchronization conditions modulo $2\pi$.

Following the asymptotic middle-state formulation of soliton synchronization~\cite{Sun2016}, the incoming and outgoing asymptotic center trajectories can be written symmetrically as
\begin{equation}
x_j^{(\pm)}(t)
=
V_jt+d_j
\mp
\frac{1}{2\beta_j}
\sum_{m\ne j}
\operatorname{sgn}(V_j-V_m)
\operatorname{Re}\Omega_{jm},
\end{equation}
with the corresponding carrier phases
\begin{equation}
\mathcal P_j^{(\pm)}(x,t)
=
\mathcal P_j(x,t)
\pm
\sum_{m\ne j}
\operatorname{sgn}(V_j-V_m)
\operatorname{Im}\Omega_{jm}.
\end{equation}
Here $(-)$ and $(+)$ denote the incoming and outgoing asymptotic states, respectively.

The position and phase corrections in these asymptotic states contain the cumulative contributions of all pairwise soliton interactions. For a single global collision event, the corresponding middle state provides a natural reference for describing the collision. Specifically, the middle trajectory and phase are obtained by averaging those of the incoming and outgoing asymptotic states:
\begin{subequations}\label{states}
\begin{equation}
\frac{x_j^{(+)}(t)+x_j^{(-)}(t)}{2}
=
V_jt+d_j,
\end{equation}
\begin{equation}
\frac{\mathcal P_j^{(+)}(x,t)+
      \mathcal P_j^{(-)}(x,t)}{2}
=
\mathcal P_j(x,t).
\end{equation}
\end{subequations}

When several collision events occur at distinct spacetime locations, however, the global asymptotic shifts combine contributions from different interactions and therefore cannot be assigned directly to an individual event. In particular, a remote soliton may shift and rephase a later collision even when it is absent from the local collision core. This motivates an eventwise construction using the pairwise interaction factors $\Omega_{jm}$ introduced above, which we develop below.

\subsection{Synchronization conditions for a collision event}

Consider a prescribed collision event $\mu$ centered at
\begin{equation}
P_\mu=(X_\mu,T_\mu),
\end{equation}
with participating solitons collected in $\mathcal C_\mu\subseteq\{1,\ldots,N\}$. The complementary set
\begin{equation}
\mathcal R_\mu
=
\{1,\ldots,N\}\setminus\mathcal C_\mu
\label{eq:remote-index-set}
\end{equation}
contains the solitons that are remote from the collision core.

For each $r\in\mathcal R_\mu$, the sign $\epsilon_r^{(\mu)}\in\{-1,+1\}$ identifies which asymptotic sector of soliton $r$ contains the event center $P_\mu$. Specifically,
\begin{equation}
\epsilon_r^{(\mu)}
=
\sgn\!\left[
\operatorname{Re}g_r(P_\mu)
\right].
\label{eq:event-sector-sign}
\end{equation}
The case $\operatorname{Re}g_r(P_\mu)=0$ corresponds to a sector boundary and is excluded by the strict sector self-consistency condition~\eqref{eq:sector-self-consistency}.
For a participating soliton $j\in\mathcal C_\mu$, the remote solitons remain well separated during the local event, so their sector signs and scattering contributions to soliton $j$ remain unchanged. In contrast, when the participating solitons have pairwise distinct velocities, their mutual scattering contributions change sign between the pre- and post-collision states. These local states can therefore be written as
\begin{equation}
\begin{aligned}
\mathcal Z_{j|\mu}^{(\pm)}(x,t)
&=2g_j(x,t)+\sum_{r\in\mathcal R_\mu}\epsilon_r^{(\mu)}\Omega_{jr}\\
&\quad\pm\sum_{\substack{m\in\mathcal C_\mu\\m\neq j}}
\operatorname{sgn}(V_j-V_m)\Omega_{jm},\quad j\in\mathcal C_\mu,
\end{aligned}
\label{eq:local-asymptotic-states}
\end{equation}
Here $(-)$ and $(+)$ denote the pre- and post-collision asymptotic states of the effective $\mathcal C_\mu$-soliton system, respectively.

For the general local formulation, we introduce the effective local state
\begin{equation}
\mathcal Z_{j|\mu}^{\mathrm{eff}}(x,t)
=
2g_j(x,t)
+
\sum_{r\in\mathcal R_\mu}
\epsilon_r^{(\mu)}\Omega_{jr},
\qquad
j\in\mathcal C_\mu .
\label{eq:local-effective-state}
\end{equation}
For pairwise distinct participating velocities, Eq.~\eqref{eq:local-effective-state} coincides with the arithmetic mean of the pre- and post-collision states in Eq.~\eqref{eq:local-asymptotic-states}. The direct definition above remains valid when some participating solitons have equal velocities.

The collision event $\mu$ is locally synchronized if, for all $j\in\mathcal C_\mu$, the corresponding effective local states are centered at the prescribed event point $P_\mu$ and share a common phase $\Phi_\mu$ modulo $2\pi$:
\begin{subequations}
\label{eq:local-sync}
\begin{align}
\operatorname{Re}
\mathcal Z_{j|\mu}^{\mathrm{eff}}(P_\mu)
&=0,
\qquad
j\in\mathcal C_\mu,
\label{eq:event-position-sync}\\
\operatorname{Im}
\mathcal Z_{j|\mu}^{\mathrm{eff}}(P_\mu)
&=
\Phi_\mu
\pmod{2\pi},
\qquad
j\in\mathcal C_\mu.
\label{eq:event-phase-sync}
\end{align}
\end{subequations}
The modular phase relation in Eq.~\eqref{eq:event-phase-sync} admits an independent integer branch $n_{j\mu}\in\mathbb Z$ for each participating pair $(j,\mu)$.

These conditions describe the synchronization of the participating solitons in the collision core. To regard $\mathcal C_\mu$ as an isolated collision event, the nonparticipating solitons must also remain sufficiently separated from the collision core. For each $r\in\mathcal R_\mu$, define the frozen-sector tail coordinate
\begin{equation}
 \mathcal Z_{r|\mu}^{\mathrm{loc}}(x,t)
 =2g_r(x,t)+\sum_{\ell\ne r}
 \epsilon_\ell^{(r|\mu)}\Omega_{r\ell}.
 \label{eq:remote-local-coordinate}
\end{equation}

Here $\epsilon_\ell^{(r|\mu)}=+1$ if $\operatorname{Re}g_\ell(P_\mu)\geq0$, and $-1$ otherwise. These signs are evaluated at $P_\mu$ and held fixed in the tail diagnostic. We require
\begin{equation}
\sech\left|
\operatorname{Re}
\mathcal Z_{r|\mu}^{\mathrm{loc}}(P_\mu)
\right|
\leq\varepsilon,
\qquad
r\in\mathcal R_\mu,
\label{eq:isolation-test}
\end{equation}
where $\varepsilon$ is a prescribed relative-tail tolerance.

Accordingly, we introduce the isolation measure
\begin{equation}
L_\mu
=
\frac{1}{2}
\min_{r\in\mathcal R_\mu}
\left|
\operatorname{Re}
\mathcal Z_{r|\mu}^{\mathrm{loc}}(P_\mu)
\right|.
\label{eq:isolation-measure}
\end{equation}
If $\mathcal R_\mu=\varnothing$, we set $L_\mu=+\infty$.

For an isolated event $P_\mu=(X_\mu,T_\mu)$ satisfying the synchronization conditions~\eqref{eq:local-sync}, let
\begin{equation}
\mathcal U_\mu
=
\{(x,t):|x-X_\mu|\le h_x,\ |t-T_\mu|\le h_t\},
\label{eq:local-neighborhood}
\end{equation}
where $h_x,h_t>0$ are fixed independently of $L_\mu$. For sufficiently large $L_\mu$, the remote sector signs remain unchanged throughout $\mathcal U_\mu$. The determinant reduction in Appendix~\ref{app:local_reduction} gives
\begin{equation}
q_N(x,t)
=
q_{\mathcal C_\mu}^{\mathrm{eff}}(x,t)
+
O\!\left(e^{-2L_\mu}\right),
\label{eq:local-reduction}
\end{equation}
uniformly on $\mathcal U_\mu$. Here $q_{\mathcal C_\mu}^{\mathrm{eff}}$ is the reduced $|\mathcal C_\mu|$-soliton solution with effective local variables
\begin{equation}
\widetilde g_{j|\mu}
=
g_j
+
\frac{1}{2}
\sum_{r\in\mathcal R_\mu}
\epsilon_r^{(\mu)}\Omega_{jr},
\qquad
j\in\mathcal C_\mu.
\label{eq:effective-local-g}
\end{equation}
Thus the finite effect of the remote solitons is retained through the scattering shifts in $\widetilde g_{j|\mu}$, while their remaining contribution is exponentially small.

At the synchronized center $P_\mu$, the effective participating states have a common phase. The center-value calculation in Appendix~\ref{app:local_reduction}, together with Eq.~\eqref{eq:local-reduction}, yields
 \begin{equation}
|q_N(P_\mu)|
=
\sum_{j\in\mathcal C_\mu}\mathcal A_j
+
O\!\left(e^{-2L_\mu}\right).
\label{eq:local-amplitude-sum}
\end{equation}
These results imply that the local Peregrine approximations reported for suitably chosen soliton collisions~\cite{Gelash2018} also apply near the corresponding isolated synchronized events in the full multisoliton field, with an additional error of $O(e^{-2L_\mu})$ in the amplitude profile.

\subsection{Inverse formulation for prescribed collision events}
\label{subsec:inverse-prescribed-events}

For prescribed event centers, we seek common norming data that satisfy the local synchronization conditions for the target collection
\begin{equation}
\mathcal E_{\mathrm{tar}}
=
\left\{
\left(
\mathcal C_\mu,
X_\mu,
T_\mu
\right)
\right\}_{\mu=1}^{K},
\label{eq:target-event-collection}
\end{equation}
where $\mathcal C_\mu$ is the participating set and $P_\mu=(X_\mu,T_\mu)$ is the prescribed center. The position condition~\eqref{eq:event-position-sync} determines the value of $s_j$ required at event $\mu$. We denote this value by $S_{j\mu}^{(x)}$:
\begin{equation}
S_{j\mu}^{(x)}
=
-2\beta_j(X_\mu-V_jT_\mu)
-
\sum_{r\notin\mathcal C_\mu}
\epsilon_r^{(\mu)}
\operatorname{Re}\Omega_{jr}.
\label{eq:required-position-norming}
\end{equation}
For a fixed soliton $j$, define
\begin{equation}
\mathcal I_j
=
\{\mu:j\in\mathcal C_\mu\},\label{eq:incident-event-set}
\end{equation}
as the set of events incident on soliton $j$. At each such event,
\begin{equation}
s_j=S_{j\mu}^{(x)}.\label{sj}
\end{equation}
Thus the position requirements are compatible if and only if
\begin{equation}
S_{j\mu}^{(x)}
=
S_{j\nu}^{(x)},
\qquad
\mu,\nu\in\mathcal I_j.\label{eq:position-compatibility}
\end{equation}
When this holds, Eq.~\eqref{sj} reconstructs $s_j$.

To express the phase conditions, define
\begin{equation}
B_{j\mu}^{(\phi)}
=
-2\alpha_jX_\mu
-4(\alpha_j^2-\beta_j^2)T_\mu
+
\sum_{r\notin\mathcal C_\mu}
\epsilon_r^{(\mu)}
\operatorname{Im}\Omega_{jr}.
\label{eq:known-event-phase}
\end{equation}
The local synchronization condition~\eqref{eq:event-phase-sync} becomes
\begin{equation}
\vartheta_j+B_{j\mu}^{(\phi)}
=
\Phi_\mu
\pmod{2\pi},
\qquad
j\in\mathcal C_\mu ,
\label{eq:phase-incidence}
\end{equation}
where $\Phi_\mu$ is the common event phase. Since each soliton has a single norming phase $\vartheta_j$, the phase relations must be consistent around every closed soliton--event sequence modulo $2\pi$. A closed sequence has the form
\begin{equation}
\mathfrak C:
\qquad
j_1-m_1-j_2-m_2-\cdots-j_p-m_p-j_1 ,
\label{eq:incidence-cycle-sequence}
\end{equation}
where $p$ denotes the number of alternating soliton--event pairs in the cycle, and
\begin{equation}
j_a,j_{a+1}\in\mathcal C_{m_a},
\qquad
a=1,\ldots,p,
\qquad
j_{p+1}=j_1 .
\label{eq:incidence-cycle-membership}
\end{equation}
The sequence records shared participation and imposes no temporal ordering. Subtracting the two phase relations at event $m_a$ gives
\begin{equation}
\vartheta_{j_{a+1}}-\vartheta_{j_a}
=
B_{j_a,m_a}^{(\phi)}
-
B_{j_{a+1},m_a}^{(\phi)}
\pmod{2\pi}.
\label{eq:phase-difference-on-event}
\end{equation}

Summing around the cycle cancels the soliton phases and yields
\begin{equation}
\sum_{a=1}^{p}
\left[
B_{j_{a+1},m_a}^{(\phi)}
-
B_{j_a,m_a}^{(\phi)}
\right]
=
2\pi n_{\mathfrak C},
\qquad
n_{\mathfrak C}\in\mathbb Z .
\label{eq:phase-cycle-compatibility}
\end{equation}
Here $n_{\mathfrak C}$ records the net integer phase branch around the sequence. A change of logarithm branches shifts the left-hand side by an integer multiple of $2\pi$, which can be absorbed by a corresponding relabeling of $n_{\mathfrak C}$. Define the set of soliton--event participations and its size by
\begin{equation}
\mathcal L
=
\{(j,\mu):j\in\mathcal C_\mu\},
\qquad
N_{\rm inc}
:=
|\mathcal L|
=
\sum_{\mu=1}^{K}|\mathcal C_\mu|.
\label{eq:incidence-edge-set}
\end{equation}

When the phase-closure conditions are satisfied, Eq.~\eqref{eq:phase-incidence} determines all phases modulo $2\pi$, up to one common phase in each connected group. The number of independent phase-closure conditions is
\begin{equation}
N_{\rm cyc}
=
N_{\rm inc}-N-K+N_{\rm comp},
\label{eq:phase-cycle-rank}
\end{equation}
where $N_{\rm comp}$ is the number of groups linked by shared solitons. If no closed sequence exists, phase reconstruction imposes no additional constraint.

The sector signs are assigned before reconstruction. With the reconstructed norming data, they must satisfy
\begin{equation}
\epsilon_r^{(\mu)}
\operatorname{Re} g_r(P_\mu)>0,
\qquad
r\notin\mathcal C_\mu .
\label{eq:sector-self-consistency}
\end{equation}
Each event must also satisfy the isolation condition~\eqref{eq:isolation-test}.

Thus, for fixed spectrum, centers, memberships, and candidate sectors, Eqs.~\eqref{eq:position-compatibility} and~\eqref{eq:phase-cycle-compatibility} on an independent set of closed sequences determine algebraic compatibility.

\section{Collision organizations, realizability, and rigidity}
\label{sec:realizability-rigidity}

In this section, we investigate how the discrete spectrum constrains the realizability of complete collision organizations and whether their relative spacetime geometry can be varied continuously at fixed eigenvalues while preserving synchronization. The general realizability and rigidity analysis also allows equal soliton velocities, provided that the discrete eigenvalues remain distinct.

\subsection{Collision organization space}
\label{subsec:physical-configuration-classification}

The target collection~\eqref{eq:target-event-collection} specifies both participating solitons and event centers. We now classify its collision organization,

\[
\mathfrak O
=
\{\mathcal C_1,\ldots,\mathcal C_K\},
\qquad
|\mathcal C_\mu|\geq2 ,
\]
while the event centers are treated as unknowns constrained by the synchronization conditions. The incidence structure of $\mathfrak O$ is denoted by $\Gamma$ and records which solitons participate in each event.

For the collision-splitting problem considered here, we restrict attention to complete collision organizations. An organization is called complete if every unordered pair of solitons is contained in exactly one synchronized collision event. For each participating set $\mathcal C_\mu$, let
\[
\mathcal P_\mu
=
\{\{i,j\}:i,j\in\mathcal C_\mu,\ i<j\}
\]
denote the set of soliton pairs contained in event $\mu$. The completeness condition requires that these pair sets form a partition of all soliton pairs:
\begin{equation}
\begin{aligned}
\bigcup_{\mu=1}^{K}\mathcal P_\mu
&=\{\{i,j\}:1\leq i<j\leq N\},\\
\mathcal P_\mu\cap\mathcal P_\nu&=\varnothing,\quad\mu\neq\nu.
\end{aligned}
\label{eq:complete-organization}
\end{equation}
Thus, different participating sets may share individual solitons, but no soliton pair can belong to more than one collision event.

This partition property also constrains the possible number of collision events and their sizes. Since an event involving $m$ solitons contains $\binom{m}{2}$ pairwise interaction channels, one has
\begin{equation}
\sum_{\mu=1}^{K}
\binom{|\mathcal C_\mu|}{2}
=
\binom{N}{2}.
\label{eq:configuration-complete-pair-count}
\end{equation}
Equivalently, starting from the fully pairwise organization with $\binom{N}{2}$ events, grouping an $m$-soliton event combines $\binom{m}{2}$ pairwise events into a single synchronized event and reduces the number of events by $ \binom{m}{2}-1 . $ Therefore,
\begin{equation}
K
=
\binom{N}{2}
-
\sum_{\mu:\,|\mathcal C_\mu|\geq3}
\left[
\binom{|\mathcal C_\mu|}{2}-1
\right].
\label{eq:configuration-complete-event-count-v2}
\end{equation}

For $N\geq3$, complete organizations can be classified into three types according to the grouping patterns of their participating sets. For $N=3$, only the collective and fully pairwise organizations are possible.

\medskip
\noindent
\textit{One-point collective collision.} All $N$ solitons synchronize at a single event,
\begin{equation}
K=1,
\qquad
\mathcal C_1=\{1,\ldots,N\},
\qquad
P_1=P_c .
\label{eq:collective-configuration-membership}
\end{equation}
The one-point collective collision is a well-studied configuration in the focusing NLS literature and serves here as the reference organization for collision splitting.

\medskip
\noindent
\textit{Partially resolved organizations.} There are multiple collision events, while at least one event still contains three or more solitons,
\begin{equation}
K>1,
\qquad
\max_{1\leq\mu\leq K}|\mathcal C_\mu|\geq3 .
\label{eq:partially-resolved-membership}
\end{equation}
Such organizations describe intermediate forms of collision splitting, in which one or more collective collision cores remain while other pairwise interaction channels are separated into distinct collision events.

\medskip
\noindent
\textit{Fully pairwise resolved organizations.} Every collision event contains exactly two solitons,
\begin{equation}
|\mathcal C_\mu|=2,
\qquad
\mu=1,\ldots,K .
\label{eq:fully-pairwise-membership}
\end{equation}
In this case, every event corresponds to a single soliton pair, and the completeness condition gives
\begin{equation}
K=\binom{N}{2}.
\label{eq:fully-pairwise-number}
\end{equation}

\subsection{Realizability of collision organizations}
\label{subsec:compatibility-matrix}

To examine how the discrete spectrum and collision organization affect realizability, we use the compatibility conditions in Sec.~\ref{subsec:inverse-prescribed-events} to constrain the event centers. For a complete collision organization, all events are linked through shared solitons, so $N_{\rm comp}=1$. We collect the event-center coordinates in
\begin{equation}
\mathbf P=(X_1,T_1,\ldots,X_K,T_K)^{\mathsf T}.
\label{eq:event-coordinate-vector}
\end{equation}

Choosing one reference event $\mu_j^0\in\mathcal I_j$ for each soliton, Eq.~\eqref{eq:position-compatibility} gives, for $\mu\in\mathcal I_j\setminus\{\mu_j^0\}$,
\begin{equation}
X_\mu+4\alpha_jT_\mu
-
X_{\mu_j^0}-4\alpha_jT_{\mu_j^0}
=
b_{j\mu}^{(x)},
\label{eq:position-compatibility-linear}
\end{equation}
where the scattering-shift difference is
\begin{align}
b_{j\mu}^{(x)}
={}&
\frac{1}{2\beta_j}
\left[
\sum_{r\notin\mathcal C_{\mu_j^0}}
\epsilon_r^{(\mu_j^0)}
\operatorname{Re}\Omega_{jr}
-
\sum_{r\notin\mathcal C_\mu}
\epsilon_r^{(\mu)}
\operatorname{Re}\Omega_{jr}
\right].
\label{eq:position-compatibility-rhs}
\end{align}
The number of position rows is
\begin{equation}
M_x
=
\sum_{j=1}^{N}(|\mathcal I_j|-1)
=
N_{\mathrm{inc}}-N .
\label{eq:number-position-rows}
\end{equation}
The number of independent phase conditions follows from Eq.~\eqref{eq:phase-cycle-rank}:
\begin{equation}
N_{\mathrm{cyc}}
=
N_{\mathrm{inc}}-N-K+1.
\label{eq:number-phase-cycles}
\end{equation}
Choose an independent set of closed sequences $\mathfrak C_1,\ldots,\mathfrak C_{N_{\mathrm{cyc}}}$.
Expanding Eq.~\eqref{eq:phase-cycle-compatibility} for the $\ell$th basis cycle, with the indexing of Eq.~\eqref{eq:incidence-cycle-sequence}, gives
\begin{equation}
\begin{aligned}
\sum_{a=1}^{p}\Big\{&2(\alpha_{j_a}-\alpha_{j_{a+1}})X_{m_a}\\
&+4\big[(\alpha_{j_a}^2-\beta_{j_a}^2)
-(\alpha_{j_{a+1}}^2-\beta_{j_{a+1}}^2)\big]T_{m_a}\Big\}\\
&=b_\ell^{(\phi,0)}+2\pi n_\ell,\qquad n_\ell\in\mathbb Z.
\end{aligned}
\label{eq:phase-compatibility-linear}
\end{equation}
The scattering contribution at zero phase branch is
\begin{equation}
\begin{aligned}
b_\ell^{(\phi,0)}
={}&\sum_{a=1}^{p}\Bigg[\sum_{r\notin\mathcal C_{m_a}}
\epsilon_r^{(m_a)}\operatorname{Im}\Omega_{j_a r}\\
&\qquad-\sum_{r\notin\mathcal C_{m_a}}
\epsilon_r^{(m_a)}\operatorname{Im}\Omega_{j_{a+1}r}\Bigg].
\end{aligned}
\label{eq:phase-compatibility-rhs}
\end{equation}
Combining the position and phase conditions yields a linear compatibility system for the event centers. The total number of equations is
\begin{equation}
M=M_x+N_{\mathrm{cyc}}
=2N_{\mathrm{inc}}-2N-K+1
\label{eq:number-compatibility-rows}
\end{equation}

Let $A_{\Gamma,x}$ contain the position rows and $A_{\Gamma,\phi}$ the phase-cycle rows. We write
\begin{equation}
A_\Gamma
=
\begin{pmatrix}
A_{\Gamma,x}\\
A_{\Gamma,\phi}
\end{pmatrix}
\in\mathbb R^{M\times2K}.
\label{eq:compatibility-matrix}
\end{equation}
With
\begin{equation}
\mathbf n
=
(n_1,\ldots,n_{N_{\mathrm{cyc}}})^{\mathsf T}
\in\mathbb Z^{N_{\mathrm{cyc}}},
\label{eq:phase-branch-vector}
\end{equation}
the right-hand side can be arranged as
\begin{equation}
\mathbf b_\Gamma(\mathbf n)
=
\mathbf b_\Gamma(\mathbf0)
+2\pi
\begin{pmatrix}
\mathbf0_{M_x}\\
\mathbf n
\end{pmatrix},
\label{eq:phase-dependent-rhs}
\end{equation}
where $\mathbf b_\Gamma(\mathbf0)$ stacks the position and zero-branch phase right-hand sides. The compatibility system is therefore
\begin{equation}
A_\Gamma(\boldsymbol\kappa)\mathbf P
=
\mathbf b_\Gamma(\boldsymbol\kappa,\boldsymbol\epsilon,\mathbf n),
\qquad
\mathbf n\in\mathbb Z^{N_{\mathrm{cyc}}}.
\label{eq:general-compatibility-system}
\end{equation}
The spectrum fixes $A_\Gamma$, while the scattering shifts, together with the remote-sector signs $\boldsymbol\epsilon$ and the integer branch $\mathbf n$, determine the right-hand side. Appendix~\ref{app:compatibility-construction} gives the explicit position and zero-branch phase right-hand sides forming $\mathbf b_\Gamma(\mathbf0)$, and hence the full $\mathbf b_\Gamma(\mathbf n)$, together with the blockwise construction of $A_{\Gamma,x}$ and $A_{\Gamma,\phi}$.

Let the columns of $L_\Gamma\in\mathbb R^{M\times\rho_\Gamma}$ form a basis of $\ker A_\Gamma^{\mathsf T}$, where
\begin{equation}
\rho_\Gamma
=\dim\ker A_\Gamma^{\mathsf T}
=M-\operatorname{rank}A_\Gamma .
\label{eq:realizability-obstruction-count}
\end{equation}
For a given organization $\Gamma$, the corresponding spectral selection condition is
\begin{equation}
F_\Gamma(\boldsymbol\kappa,\boldsymbol\epsilon,\mathbf n)
:=
L_\Gamma^{\mathsf T}
\mathbf b_\Gamma(\boldsymbol\kappa,\boldsymbol\epsilon,\mathbf n)
=0 .
\label{eq:spectral-selection-rule}
\end{equation}

\begin{proposition}[Realizability of the reduced synchronization system]
\label{prop:reduced-realizability}
For a prescribed organization $\Gamma$ and fixed spectrum, sector assignment, and phase branches, there exist event centers and common norming data satisfying all local synchronization equations of $\Gamma$ if and only if Eq.~\eqref{eq:spectral-selection-rule} holds.
\end{proposition}

\noindent\textit{Proof.}
Consider the compatibility system \eqref{eq:general-compatibility-system}. Since the columns of $L_\Gamma$ form a basis of $\ker A_\Gamma^{\mathsf T}$,
\[
F_\Gamma=0
\quad\Longleftrightarrow\quad
\mathbf b_\Gamma\perp\ker A_\Gamma^{\mathsf T}
\quad\Longleftrightarrow\quad
\mathbf b_\Gamma\in\operatorname{Col}A_\Gamma .
\]
Hence Eq.~\eqref{eq:general-compatibility-system} admits a solution for the event centers. The reconstruction in Sec.~\ref{subsec:inverse-prescribed-events} then yields common norming data, up to one global phase. Conversely, any synchronized realization satisfies Eq.~\eqref{eq:general-compatibility-system} and hence $F_\Gamma=0$.
\hfill$\square$

The spectrum supports the organization algebraically if and only if the selection rule holds for at least one sector assignment and integer phase branch $(\boldsymbol\epsilon,\mathbf n)$.

The algebraic compatibility conditions characterize the candidate collision geometries. Physical admissibility further requires the sector self-consistency condition~\eqref{eq:sector-self-consistency}, the isolation condition~\eqref{eq:isolation-test}, and separation of distinct event centers. When free event coordinates remain, these requirements determine the physically admissible portion of the compatible solution family.

\subsection{Spectral rigidity and compatibility balance}
\label{subsec:spectral-hyperframework}

At fixed spectrum, sector assignment, and phase branch, two compatible geometries $\mathbf P$ and $\mathbf P'$ satisfy Eq.~\eqref{eq:general-compatibility-system}. Their difference $\delta\mathbf P=\mathbf P'-\mathbf P$ therefore satisfies
\begin{equation}
A_\Gamma\delta\mathbf P=0.
\label{eq:homogeneous-event-compatibility}
\end{equation}

The kernel $\ker A_\Gamma$ contains a two-dimensional trivial subspace generated by uniform translations of all event centers in space and time. Factoring out this subspace defines the internal deformation dimension,
\begin{equation}
D_{\mathrm{int}}(\boldsymbol\kappa,\Gamma)
=\dim\ker A_\Gamma-2.
\label{eq:internal-dimension-main}
\end{equation}
Thus $D_{\mathrm{int}}=0$ means that the relative event geometry is rigid, whereas $D_{\mathrm{int}}>0$ allows nontrivial algebraic internal deformations.

For $\delta\mathbf P\in\ker A_\Gamma$, the homogeneous position conditions~\eqref{eq:position-compatibility-linear} make $\delta X_\mu+4\alpha_j\delta T_\mu$ independent of the event containing soliton $j$. Denote this common value by

\begin{equation}
\xi_j:=\delta X_\mu+4\alpha_j\delta T_\mu
=-\frac{\delta s_j}{2\beta_j},
\qquad j\in\mathcal C_\mu.
\label{eq:xi-incidence}
\end{equation}
The last equality follows by varying the position synchronization condition~\eqref{eq:event-position-sync}.

Varying the phase synchronization condition~\eqref{eq:event-phase-sync} gives
\begin{equation}
\delta\vartheta_j
=
\delta\Phi_\mu
+2\alpha_j\delta X_\mu
+4(\alpha_j^2-\beta_j^2)\delta T_\mu,
\qquad
j\in\mathcal C_\mu.
\label{eq:eta-incidence}
\end{equation}
Since $\delta\mathbf P\in\ker A_\Gamma$, the phase variations can be reconstructed consistently, up to a common additive constant, as shown in Appendix~\ref{app:event-spectral-equivalence}. Together with $\delta s_j=-2\beta_j\xi_j$, they determine the norming data variations.

For each eigenvalue $\kappa_j$, define the spectral point
\begin{equation}
\mathbf p_j=\bigl(|\kappa_j|^2,\alpha_j\bigr)^{\mathsf T},
\label{eq:spectral-point}
\end{equation}
and let
\begin{equation}
J=
\begin{pmatrix}
0&1\\
-1&0
\end{pmatrix}.
\label{eq:spectral-J}
\end{equation}
Introduce
\begin{equation}
\begin{aligned}
\mathbf u_j&=\begin{pmatrix}\xi_j\\\delta\vartheta_j-2\alpha_j\xi_j\end{pmatrix},\\
\mathbf a_\mu&=\begin{pmatrix}\delta X_\mu\\\delta\Phi_\mu\end{pmatrix},
\qquad\omega_\mu=4\delta T_\mu.
\end{aligned}
\label{eq:spectral-deformation-variable}
\end{equation}
Equations~\eqref{eq:xi-incidence} and \eqref{eq:eta-incidence} then give
\begin{equation}
\mathbf u_j=\mathbf a_\mu+\omega_\mu J\mathbf p_j,
\qquad j\in\mathcal C_\mu.
\label{eq:event-rigid-motion}
\end{equation}
These auxiliary motions represent norming data variations at fixed eigenvalues. For each event, Eq.~\eqref{eq:event-rigid-motion} describes a common infinitesimal rigid motion of the participating auxiliary spectral points, with two translational parameters and one rotational parameter. These constraints can be represented by any full-row-rank matrix $R_\mu(\mathbf p)$ satisfying
\begin{equation}
\begin{aligned}
\ker R_\mu(\mathbf p)=\bigl\{&
(\mathbf u_j)_{j\in\mathcal C_\mu}:
\exists\,\mathbf a_\mu\in\mathbb R^2,\
\omega_\mu\in\mathbb R,\\
&\mathbf u_j=\mathbf a_\mu+\omega_\mu J\mathbf p_j
\quad\text{for all }j\in\mathcal C_\mu\bigr\}.
\end{aligned}
\label{eq:local-rigidity-nullspace}
\end{equation}
Since each event contains at least two distinct spectral points,
\begin{equation}
\operatorname{rank}R_\mu
=
2|\mathcal C_\mu|-3.
\label{eq:local-rigidity-rank}
\end{equation}

For a two-soliton event, Eq.~\eqref{eq:event-rigid-motion} implies
\begin{equation}
(\mathbf p_i-\mathbf p_j)^{\mathsf T}
(\mathbf u_i-\mathbf u_j)=0.
\label{eq:pairwise-bar-constraint}
\end{equation}
This is the usual infinitesimal bar constraint for the pair $(\mathbf p_i,\mathbf p_j)$. For collective events, however, $R_\mu$ enforces the common rigid-motion condition \eqref{eq:event-rigid-motion} for all participating solitons.

Embedding the local constraints in
\[
\mathbf u
=
(\mathbf u_1^{\mathsf T},\ldots,
\mathbf u_N^{\mathsf T})^{\mathsf T}
\in\mathbb R^{2N}
\]
and stacking them over all prescribed events gives the global rigidity system
\begin{equation}
R_\Gamma(\mathbf p)\mathbf u=0.
\label{eq:spectral-rigidity-system}
\end{equation}
Here each shared soliton has one vector $\mathbf u_j$ across all its events. Global planar translations and rotations lie in $\ker R_\Gamma$, so $\dim\ker R_\Gamma\ge3$. The following theorem identifies the event and spectral deformation spaces after removing their trivial motions.

\begin{theorem}[Spectral rigidity representation]
\label{thm:rigidity-balance}

For a complete organization with distinct eigenvalues and an algebraically realizable branch, let $\mathcal T_A\subseteq\ker A_\Gamma$ be the two-dimensional space of uniform event translations
\[
\delta X_\mu=a,
\qquad
\delta T_\mu=b
\qquad
\text{for all }\mu,
\]
with common parameters $a,b\in\mathbb R$. Let $\mathcal T_R\subseteq\ker R_\Gamma$ be the three-dimensional space of global infinitesimal rigid motions of the auxiliary spectral points,
\[
\mathbf u_j
=
\mathbf a+\omega J\mathbf p_j
\qquad
\text{for all }j,
\]
with common parameters $\mathbf a\in\mathbb R^2$ and $\omega\in\mathbb R$.

The correspondence map $\Pi:\ker R_\Gamma\to\ker A_\Gamma$ constructed in Appendix~\ref{app:event-spectral-equivalence} induces a linear isomorphism
\begin{equation}
\overline{\Pi}:
\ker R_\Gamma/\mathcal T_R
\xrightarrow{\;\cong\;}
\ker A_\Gamma/\mathcal T_A,
\qquad
[\mathbf u]\longmapsto[\Pi(\mathbf u)].
\label{eq:quotient-deformation-correspondence-main}
\end{equation}
Here $[\mathbf u]$ and $[\Pi(\mathbf u)]$ denote equivalence classes modulo $\mathcal T_R$ and $\mathcal T_A$, respectively. Thus, after the respective trivial motions are removed, auxiliary spectral motions and event deformations are in one-to-one linear correspondence.

Consequently,
\begin{equation}
D_{\mathrm{int}}
=
\dim\ker R_\Gamma-3
=
2N-3-\operatorname{rank}R_\Gamma.
\label{eq:Dint-spectral-rigidity}
\end{equation}
Hence the relative event geometry is rigid if and only if
\begin{equation}
\operatorname{rank}R_\Gamma=2N-3.
\label{eq:spectral-rigidity-criterion}
\end{equation}

\end{theorem}

\noindent\textit{Proof.}
Appendix~\ref{app:event-spectral-equivalence} proves the quotient space isomorphism. Taking dimensions and using $\dim\mathcal T_A=2$ and $\dim\mathcal T_R=3$ gives $D_{\mathrm{int}}=\dim\ker R_\Gamma-3$. Since $R_\Gamma$ has $2N$ columns, the rank--nullity theorem yields Eq.~\eqref{eq:Dint-spectral-rigidity}. The rigidity criterion~\eqref{eq:spectral-rigidity-criterion} follows by setting $D_{\mathrm{int}}=0$ in Eq.~\eqref{eq:Dint-spectral-rigidity}.
\hfill$\square$

\begin{corollary}[Rigidity for noncollinear spectra]
\label{cor:complete-rigidity}
For a complete organization, suppose that the spectral points $\mathbf p_j=(|\kappa_j|^2,\alpha_j)^{\mathsf T}$ are not all collinear. Then, on every algebraically realizable branch, the compatible event centers admit only uniform translations in space and time. Consequently,
\begin{equation}
D_{\mathrm{int}}=0.
\end{equation}
\end{corollary}

\noindent\textit{Proof.}
For participants in the same event, Eq.~\eqref{eq:event-rigid-motion} implies the pairwise constraint~\eqref{eq:pairwise-bar-constraint}. Since the organization is complete, this constraint holds for every pair of spectral points.

Choose three noncollinear spectral points, labeled $1$, $2$, and $3$. The pairwise constraints on these points allow only a common infinitesimal planar translation and rotation. Thus there exist $\mathbf a\in\mathbb R^2$ and $\omega\in\mathbb R$ such that
\[
\mathbf u_i=\mathbf a+\omega J\mathbf p_i,
\qquad i=1,2,3.
\]
Subtract this global rigid motion and write
\[
\widetilde{\mathbf u}_j
=\mathbf u_j-\mathbf a-\omega J\mathbf p_j.
\]
The residual vectors still satisfy all pairwise constraints, and $\widetilde{\mathbf u}_i=0$ for $i=1,2,3$.

For any remaining spectral point $\mathbf p_j$, the constraints with the three chosen points give
\[
(\mathbf p_j-\mathbf p_i)^{\mathsf T}
\widetilde{\mathbf u}_j=0,
\qquad i=1,2,3.
\]
At least two of the vectors $\mathbf p_j-\mathbf p_i$ are linearly independent, since the three chosen points are noncollinear. Hence $\widetilde{\mathbf u}_j=0$.

Therefore every $\mathbf u\in\ker R_\Gamma$ is a global infinitesimal planar rigid motion. Since all such motions belong to $\ker R_\Gamma$,
\[
\ker R_\Gamma=\mathcal T_R,
\qquad
\dim\ker R_\Gamma=3.
\]
Theorem~\ref{thm:rigidity-balance} then gives $D_{\mathrm{int}}=0$.
\hfill$\square$

With the discrete spectrum and complete collision organization fixed, $A_\Gamma$ is fixed. Changes in the sector assignment and integer phase branch allow the right-hand side to take only discrete values. During a continuous deformation of the event centers, $A_\Gamma\mathbf P$ is continuous, so the right-hand side must remain constant. The difference between the event-center vectors before and after such a deformation therefore lies in $\ker A_\Gamma$. By the preceding rigidity result, $\ker A_\Gamma=\mathcal T_A$, so any such deformation is restricted to a common spacetime translation of all event centers.

For later use, define
\begin{equation}
\mathcal D_{ijk}
=
\det
\begin{pmatrix}
1&|\kappa_i|^2&\alpha_i\\
1&|\kappa_j|^2&\alpha_j\\
1&|\kappa_k|^2&\alpha_k
\end{pmatrix}.
\label{eq:spectral-area-determinant}
\end{equation}
Then $\mathcal D_{ijk}=0$ if and only if the three spectral points $(|\kappa_i|^2,\alpha_i)$, $(|\kappa_j|^2,\alpha_j)$, and $(|\kappa_k|^2,\alpha_k)$ are collinear. Hence Corollary~\ref{cor:complete-rigidity} implies
\begin{equation}
D_{\mathrm{int}}>0
\quad\Longrightarrow\quad
\mathcal D_{ijk}=0
\qquad
\text{for every triple }(i,j,k).
\end{equation}
Collinearity is therefore necessary but not sufficient for internal freedom: collective-event constraints can still enforce rigidity. Algebraic realizability also requires the spectral selection conditions.

\paragraph{Compatibility balance.}
Combining Eqs.~\eqref{eq:internal-dimension-main} and \eqref{eq:realizability-obstruction-count} gives
\begin{equation}
\rho_\Gamma=\Delta_\Gamma+D_{\mathrm{int}},
\label{eq:realizability-rigidity-identity}
\end{equation}
where the constraint surplus
\begin{equation}
\Delta_\Gamma:=M-2K+2
=2N_{\mathrm{inc}}-2N-3K+3
\label{eq:constraint-surplus}
\end{equation}
depends only on the organization. At fixed organization, a rank loss increases $D_{\mathrm{int}}$ and the number $\rho_\Gamma$ of independent linear conditions on $\mathbf b_\Gamma$ by the same amount. For an algebraically realizable rigid branch, $D_{\mathrm{int}}=0$, and hence $ \rho_\Gamma=\Delta_\Gamma. $

\paragraph{Explicit consequences for two representative families.}

\begin{corollary}[Fully pairwise organizations]
\label{cor:pairwise-rigidity}
For $N\geq3$ with all $\binom{N}{2}$ pair events prescribed, the internal deformation dimension is determined by the collinearity of the auxiliary spectral points $\mathbf p_j$. If the points are not all collinear, then
\begin{equation}
D_{\mathrm{int}}=0,
\qquad
\rho_\Gamma
=
\frac{(N-2)(N-3)}{2}.
\label{eq:complete-pairwise-noncollinear-nullities}
\end{equation}
If the points are all collinear, then
\begin{equation}
D_{\mathrm{int}}=N-2,
\qquad
\rho_\Gamma
=
\frac{(N-1)(N-2)}{2}.
\label{eq:complete-pairwise-collinear-nullities}
\end{equation}
\end{corollary}

\begin{corollary}[One collective event with pairwise remainder]
\label{cor:mixed-collective-pairwise}
Consider a complete organization consisting of one $m$-soliton collective event and all remaining interactions pairwise, with $3\leq m\leq N$. If the auxiliary spectral points are not all collinear, then
\begin{equation}
D_{\mathrm{int}}=0,
\qquad
\rho_\Gamma
=
\frac{(N-m)(N+m-5)}{2}.
\label{eq:mixed-noncollinear-nullities}
\end{equation}
If the auxiliary spectral points are all collinear, then
\begin{equation}
D_{\mathrm{int}}=N-m,
\qquad
\rho_\Gamma
=
\frac{(N-m)(N+m-3)}{2}.
\label{eq:mixed-collinear-nullities}
\end{equation}
\end{corollary}

The proofs of Corollaries~\ref{cor:pairwise-rigidity} and~\ref{cor:mixed-collective-pairwise} are given in Appendix~\ref{app:explicit-rigidity-counts}. On algebraically realizable branches, the corresponding values of $D_{\mathrm{int}}$ give the dimensions of the compatible relative event geometries.

On the collinear locus, replacing the pairwise interactions among $m$ solitons by a single collective event reduces the internal deformation dimension by $m-2$. In particular, the endpoint $m=N$ is rigid even when all auxiliary spectral points are collinear. For a general complete organization, the internal deformation dimension is determined by $\operatorname{rank}R_\Gamma$ through Eq.~\eqref{eq:Dint-spectral-rigidity}, while the compatibility count follows from Eq.~\eqref{eq:realizability-rigidity-identity}.

\subsection{Examples of spectral selection and deformation}
\label{subsec:explicit-spectral-consequences}
\label{sec:fixed-spectrum-unfolding}

The three- and four-soliton examples below illustrate these rigidity alternatives and identify the spectral selection conditions.

\subsubsection{Three-soliton organizations}
\label{subsec:three-soliton-fixed-spectrum}

For $N=3$, the only complete organizations are the collective event $\{123\}$ and the fully pairwise organization $\{12,13,23\}$, as discussed in Sec.~\ref{subsec:physical-configuration-classification}. The collective organization has $(D_{\mathrm{int}},\rho_\Gamma)=(0,0)$, whereas Corollary~\ref{cor:pairwise-rigidity} gives 
\begin{equation} 
(D_{\mathrm{int}},\rho_\Gamma)
=
\begin{cases}
(0,0), & \mathcal D_{123}\ne0,\\[1mm]
(1,1), & \mathcal D_{123}=0,
\end{cases}
\label{eq:three-soliton-spectral-rigidity}
\end{equation}
for the fully pairwise organization.

For noncollinear auxiliary spectral points, this organization is algebraically realizable and rigid. If the soliton velocities are also distinct, the compatibility equations show that $|\operatorname{Re}g_r(P_\mu)|$ grows without bound and the event centers move apart as the integer phase branch $n$ tends to $+\infty$. A suitable choice of sector signs then gives a physically admissible realization of the fully pairwise organization for sufficiently large positive $n$.

For collinear points, a one-parameter family of relative event geometries exists only when the compatibility condition is satisfied.

We set
\[
\epsilon_{12}:=\epsilon_3^{(12)},\quad
\epsilon_{13}:=\epsilon_2^{(13)},\quad
\epsilon_{23}:=\epsilon_1^{(23)},
\]
where $\epsilon_{12},\epsilon_{13},\epsilon_{23}\in\{-1,+1\}$.

\paragraph{Rigid noncollinear realization.}

Consider the fixed spectrum
\begin{equation}
\kappa_1=-\frac14+\frac12\ii,
\qquad
\kappa_2=\frac12\ii,
\qquad
\kappa_3=\frac14+\frac12\ii.
\label{eq:three-soliton-spectrum}
\end{equation}
For this spectrum,
\[
\mathcal D_{123}=-\frac1{32}\ne0,
\]
so the fully pairwise organization is rigid with $(D_{\mathrm{int}},\rho_\Gamma)=(0,0)$.

\begin{figure*}[t]
  \centering
  \includegraphics[width=\linewidth]
{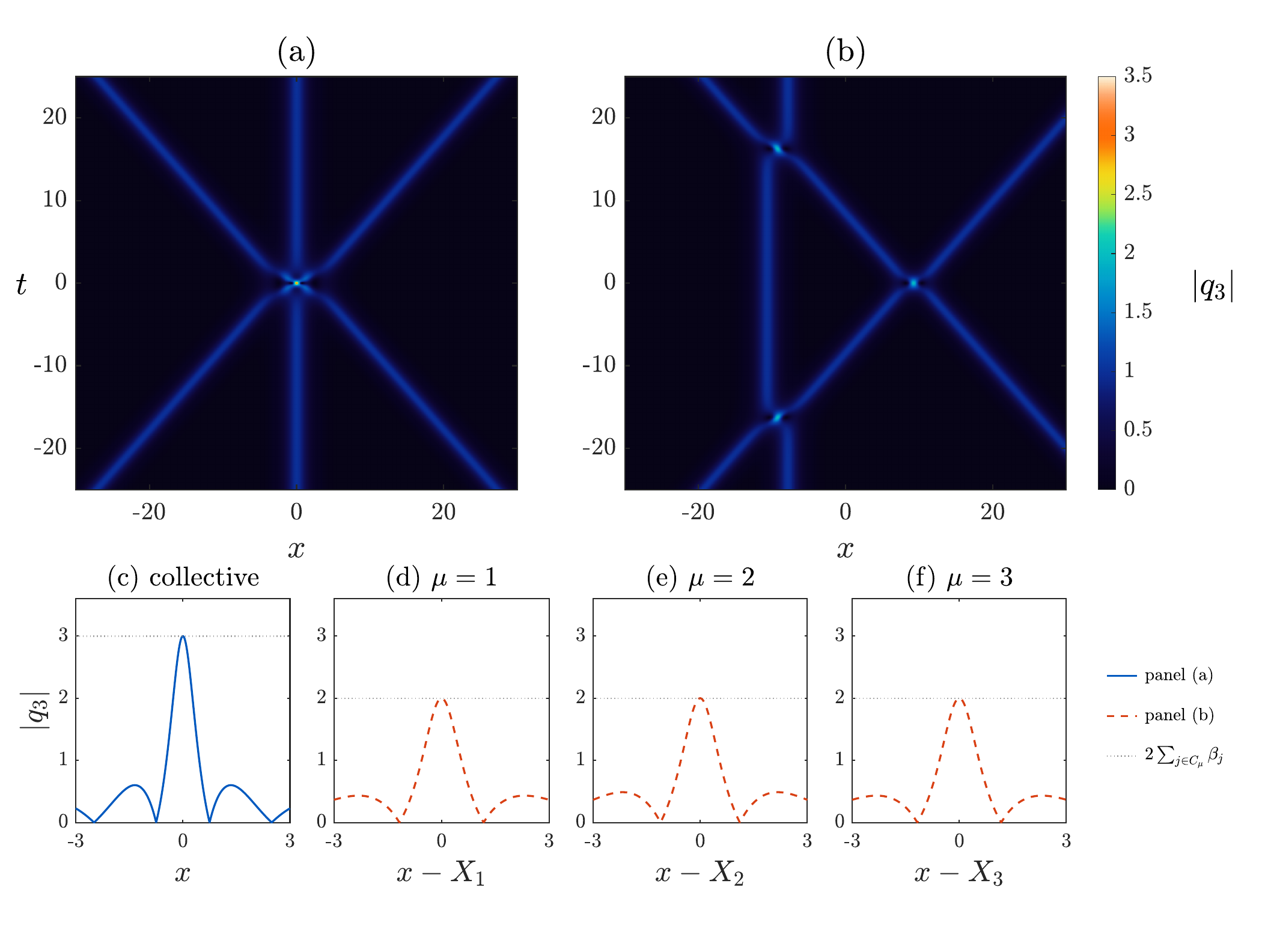} \caption{Three-soliton collision organizations for the fixed spectrum~\eqref{eq:three-soliton-spectrum}. The collective realization uses $(c_1,c_2,c_3)=(0,0,0)$. For the fully pairwise realization, $c_1=\frac12\ln17-\Delta-\frac{\ii}{2}\Delta$, $c_2=\Delta-\frac12\ln17+\ii(\arctan4-2\pi)$, and $c_3=\frac12\ln17-\Delta+ \ii(2\arctan4+\frac12\Delta)$, where $\Delta=4\pi-\frac14\ln85-2\arctan2$. The synchronized centers are $P_{12}=(-\Delta,-\tau)$, $P_{13}=(\Delta,0)$, and $P_{23}=(-\Delta,\tau)$, with $\tau=2\Delta-\frac12\ln85$. The upper panels show the collective and fully pairwise fields, and the lower panels present retarded-time profiles at the propagation distances of the corresponding collisions. Dotted gray lines indicate the synchronized amplitude sums.}
  \label{fig:three-split-map-new}
\end{figure*}

At this same spectrum, both the collective organization and a fully pairwise organization are realizable through different choices of norming data. For the pairwise realization,
\[
(\epsilon_{12},\epsilon_{13},\epsilon_{23})
=
(-1,+1,-1),
\]
and direct substitution verifies the position and phase synchronization conditions at three distinct pairwise event centers. Moreover,
\[
\max_{\mu,r}
\sech\left|
\operatorname{Re}
\mathcal Z_{r|\mu}^{\mathrm{loc}}(P_\mu)
\right|
<1.32\times10^{-6},
\]
so the prescribed sector and isolation conditions are satisfied. The two realizations are compared in Fig.~\ref{fig:three-split-map-new}.

Thus the same discrete spectrum supports two distinct complete organizations through different norming data, while the fully pairwise realization remains rigid within its fixed sector and phase branch.

\paragraph{Flexible collinear realization.}

To construct a flexible realization, let $r>0$ and set
\begin{equation}
(z_1,z_2,z_3)
=
\left(\frac r2,r,2r\right),
\qquad
\kappa_j
=
\frac{z_j+\ii}{z_j-\ii},
\qquad
j=1,2,3.
\label{eq:closed-soft-spectrum}
\end{equation}
Since $|\kappa_j|=1$, the auxiliary spectral points $\mathbf p_j=(1,\alpha_j)^{\mathsf T}$ are collinear, and hence Eq.~\eqref{eq:three-soliton-spectral-rigidity} gives $D_{\mathrm{int}}=1$ and $\rho_\Gamma=1$. The geometric progression of the $z_j$ also gives $\operatorname{Re}\Omega_{12} =\operatorname{Re}\Omega_{23}=-\ln3$ and $\operatorname{Re}\Omega_{13}=-\ln(5/3)$.

Consider the candidate sector assignment $(\epsilon_{12},\epsilon_{13},\epsilon_{23})=(-1,+1,-1)$. Fix $P_{13}=(0,0)$ to remove the two global spacetime translations, and choose the global phase gauge $\vartheta_1=0$. Since $D_{\mathrm{int}}=1$, one real relative degree of freedom remains, which we parameterize by
\begin{equation}
X_{12}+4\alpha_2T_{12}
=
X_{23}+4\alpha_2T_{23}
=
-\frac D2.
\label{eq:closed-soft-D-param}
\end{equation}
Using the position reconstruction $s_j=S_{j\mu}^{(x)}$ and the phase-incidence relation~\eqref{eq:phase-incidence}, one obtains
\begin{equation}
\begin{aligned}
c_1&=\ln3,\\
c_2&=-\ln3-\ii\kappa_2D+\ii w_1\ln5,\\
c_3&=\ln3+\ii\pi,\\
w_j&=\frac{\alpha_j}{\beta_j}=\frac{z_j-z_j^{-1}}{2}.
\end{aligned}
\label{eq:closed-soft-norming}
\end{equation}
The remaining position equations determine $P_{12}$ and $P_{23}$ affinely in $D$.

For these reconstructed event centers, the local phase synchronization conditions are automatically satisfied at $P_{12}$ and $P_{13}$. The remaining phase mismatch at $P_{23}$ is $(w_1-w_3)\ln5-\pi$, which must equal $2\pi n$, with $n\in\mathbb Z$. Choosing $n=-1$ and using $w_1-w_3=-\tfrac34(r+r^{-1})$ gives $r+r^{-1}=4\pi/(3\ln5)$. We therefore take the smaller positive root
\begin{equation}
r=
\frac{3\ln5}
{2\pi+\sqrt{4\pi^2-9\ln^2 5}}.
\label{eq:closed-soft-r}
\end{equation}
For this choice, all three pair events satisfy the local phase synchronization conditions modulo $2\pi$.

With $r$ now fixed, the discrete spectrum, sector assignment, and integer phase branch remain unchanged as $D$ varies, while the relative event geometry moves along the one-dimensional compatible family.

\begin{figure*}[t]
  \centering
  \includegraphics[width=\linewidth]
{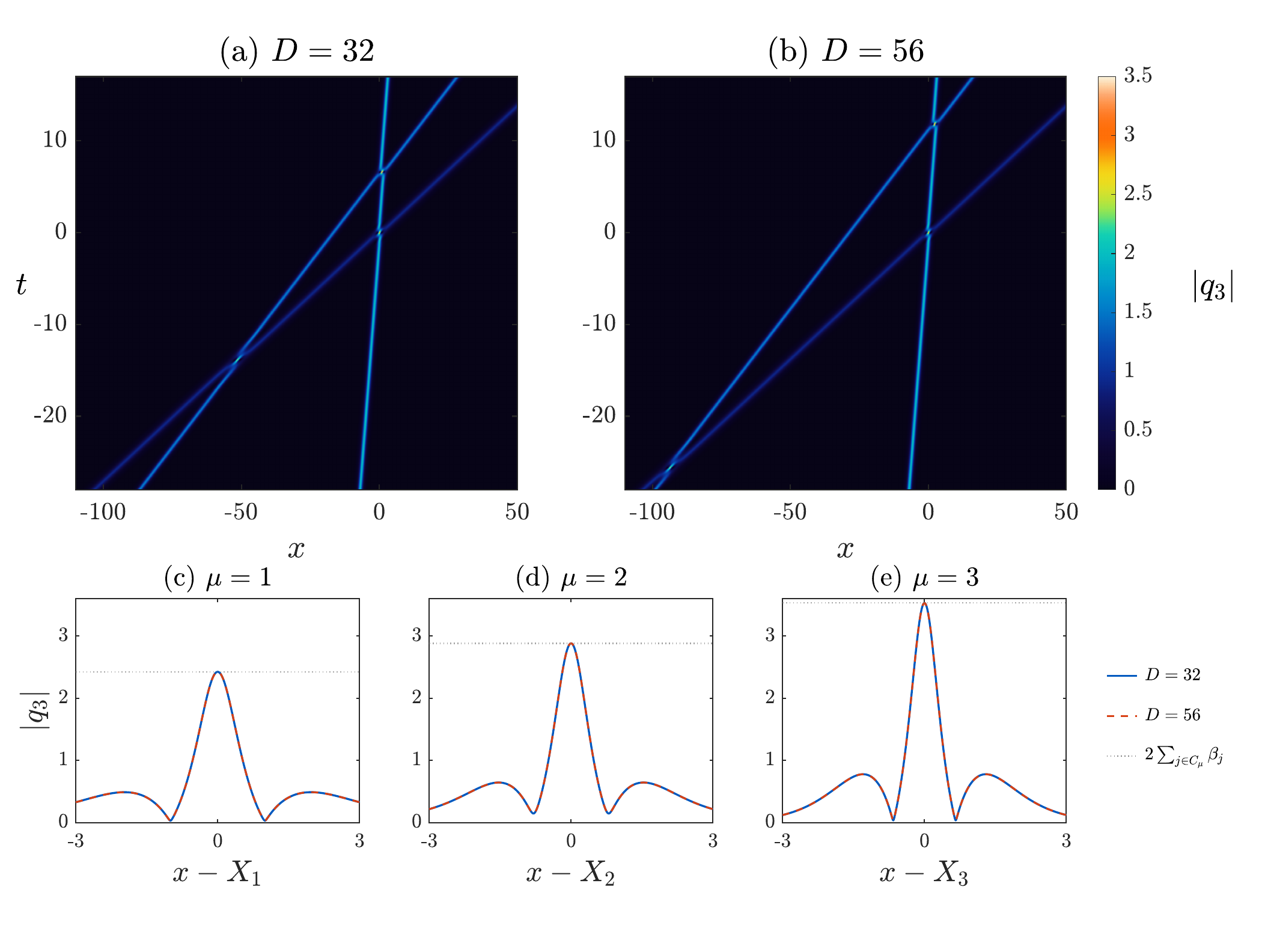}  \caption{Two members of the flexible three-soliton family for $D=32$ and $D=56$, using the fixed spectrum from Eqs.~\eqref{eq:closed-soft-spectrum} and~\eqref{eq:closed-soft-r} and the norming data in Eq.~\eqref{eq:closed-soft-norming}. For the plotted realizations, $P_{13}=(0,0)$, $T_{12}=(A-D/2)/[4(\alpha_2-\alpha_1)]$, $X_{12}=-D/2-4\alpha_2T_{12}$, $T_{23}=(B-D/2)/[4(\alpha_2-\alpha_3)]$, and $X_{23}=-D/2-4\alpha_2T_{23}$, where $A=\ln5/(2\beta_1)$ and $B=\ln5/(2\beta_3)$. As $D$ varies, the relative geometry of the pairwise collisions changes while the sector assignment and phase branch remain fixed. The lower panels show retarded-time profiles at the propagation distances of the corresponding collisions. Dotted gray lines indicate the synchronized-pair amplitude sums.}
  \label{fig:closed-soft-field}
\end{figure*}

We restrict attention to the subinterval $32\leq D\leq56$. On this fixed branch, the sector margins, the real remote local coordinates entering the isolation test, and the relevant event-time differences are affine functions of $D$ and retain their signs throughout this interval. Hence the endpoint checks suffice to verify admissibility on the whole interval. They recover the prescribed sector assignment and give
\[
\begin{aligned}
&T_{12}<T_{13}=0<T_{23},\\
&\max_{\mu,r}\sech\left|\operatorname{Re}
\mathcal Z_{r|\mu}^{\mathrm{loc}}(P_\mu)\right|<5.0\times10^{-8}.
\end{aligned}
\]
Thus the whole subinterval is physically admissible. Varying $D$ within it realizes the one-dimensional internal freedom predicted by Eq.~\eqref{eq:three-soliton-spectral-rigidity}. Figure~\ref{fig:closed-soft-field} shows the two endpoint members $D=32$ and $D=56$.

\paragraph{A spectrum with collinear auxiliary points excluded by compatibility.}

Collinearity alone does not guarantee realizability of the fully pairwise organization. Consider
\begin{equation}
\kappa_1=-\frac12+\frac{\sqrt3}{2}\ii,
\qquad
\kappa_2=\ii,
\qquad
\kappa_3=\frac12+\frac{\sqrt3}{2}\ii.
\label{eq:three-forbidden-spectrum}
\end{equation}
Since $|\kappa_j|=1$, the auxiliary spectral points $\mathbf p_j=(1,\alpha_j)^{\mathsf T}$ are collinear, and hence Eq.~\eqref{eq:three-soliton-spectral-rigidity} gives $D_{\mathrm{int}}=1$ and $\rho_\Gamma=1$. Thus one nontrivial compatibility condition must be satisfied.

For the oriented cycle
\[
1-12-2-23-3-13-1,
\]
using the construction in Appendix~\ref{app:compatibility-construction}, the compatibility matrix has a one-dimensional left null space generated by
\[
\ell=(-1,0,1,1)^{\mathsf T}.
\]
The selection condition $\ell^{\mathsf T}\mathbf b_\Gamma(n)=0$ therefore reduces to
\begin{equation}
\begin{aligned}
n&=\frac{\epsilon_{13}}{2}
+\frac{2\epsilon_{13}L-(\epsilon_{12}+\epsilon_{23})\ln2}{2\pi\sqrt3},\\
&L=\ln(2+\sqrt3),\qquad n\in\mathbb Z,
\end{aligned}
\label{eq:three-forbidden-selection}
\end{equation}
where the principal logarithm in Eq.~\eqref{eq:scattering-log-principal} is used.

Since $L>\ln2$, the second term on the right-hand side has the same sign as $\epsilon_{13}$. Moreover, for every sector assignment,
\begin{equation}
0<
\left|
\frac{
2\epsilon_{13}L
-(\epsilon_{12}+\epsilon_{23})\ln2
}{
2\pi\sqrt3
}
\right|
\leq
\frac{L+\ln2}{\pi\sqrt3}
<
\frac12.
\label{eq:three-forbidden-branch-bound}
\end{equation}
Consequently,
\[
\begin{aligned}
\frac12<n<1&\quad\text{if }\epsilon_{13}=+1,\\
-1<n<-\frac12&\quad\text{if }\epsilon_{13}=-1,
\end{aligned}
\]
which is incompatible with $n\in\mathbb Z$. Thus no sector assignment or integer phase branch realizes the fully pairwise organization at \eqref{eq:three-forbidden-spectrum}, although the same spectrum supports the collective collision with $(c_1,c_2,c_3)=(0,0,0)$.

For the three-soliton pairwise organization, these examples distinguish the two roles of the spectral conditions: collinearity determines whether an internal deformation mode is present, whereas the compatibility condition~\eqref{eq:spectral-selection-rule} determines whether the pairwise branch is realizable.

\subsubsection{Four-soliton organizations}
\label{subsec:four-soliton-fixed-spectrum}

By the classification in Sec.~\ref{subsec:physical-configuration-classification}, up to relabeling the complete four-soliton organizations are
\begin{align}
\Gamma_{\mathrm c}&=\{1234\},\nonumber\\
\Gamma_{\mathrm m}&=\{123,14,24,34\},\nonumber\\
\Gamma_{\mathrm p}&=\{12,13,14,23,24,34\}.
\label{eq:four-complete-organizations}
\end{align}
We apply the general compatibility and rigidity results above to these three organizations, assuming pairwise distinct velocities.

\paragraph{Collective collision.}
For $\Gamma_{\mathrm c}$ there are no position compatibility rows or phase cycles. Hence the selection condition is vacuous, and Corollary~\ref{cor:mixed-collective-pairwise} at $m=N=4$ gives $(D_{\mathrm{int}},\rho_\Gamma)=(0,0)$. Thus the collective organization is algebraically realizable for every spectrum in the stated regime.

\paragraph{Partially resolved collision.}
Consider the mixed organization $\Gamma_{\mathrm m}=\{123,14,24,34\}$. For $N=4$ and $m=3$, Corollary~\ref{cor:mixed-collective-pairwise} gives $(D_{\mathrm{int}},\rho_{\Gamma_{\mathrm m}})=(0,1)$ for noncollinear auxiliary spectral points and $(1,2)$ when the auxiliary spectral points are collinear. It remains to make the corresponding compatibility conditions explicit.

Fix $P_{123}=(0,0)$ and collect the remaining event centers in $\mathbf P_{\mathrm m} =(X_{14},T_{14},X_{24},T_{24},X_{34},T_{34})^{\mathsf T}$. Let $\widehat A_{\mathrm m,x}$ and $\widehat A_{\mathrm m,\phi}$ denote the position and phase blocks of the general compatibility system \eqref{eq:general-compatibility-system} after this gauge fixing.

For distinct velocities, $\operatorname{rank}\widehat A_{\mathrm m,x}=5$, so the position-compatible geometries form a one-dimensional affine family. Let $\mathbf P_{\mathrm m,0}$ be the unique solution with $T_{14}=0$, and let $\mathbf v$ span $\ker\widehat A_{\mathrm m,x}$, normalized by $v_2=1$. Then every position-compatible geometry has the form
\begin{equation}
\mathbf P_{\mathrm m}
=
\mathbf P_{\mathrm m,0}
+
\mathbf vT,
\qquad
T=T_{14}.
\label{eq:3222-position-family}
\end{equation}

There are two independent phase cycles, which we choose as $124$ and $134$. Write $\mathbf b_{\mathrm m,\phi}(\mathbf n) =(\Psi_{124},\Psi_{134})^{\mathsf T}$, where $\Psi_{abc}=b_{abc}^{(\phi,0)}+2\pi n_{abc}$, and define $(Q_{124},Q_{134})^{\mathsf T} :=\widehat A_{\mathrm m,\phi}\mathbf P_{\mathrm m,0}$. Restricting the phase compatibility equations to Eq.~\eqref{eq:3222-position-family} gives
\begin{equation}
\begin{aligned}
G_{\mathrm m}T&=\mathbf g_{\mathrm m},\\
G_{\mathrm m}&=\widehat A_{\mathrm m,\phi}\mathbf v
=4\begin{pmatrix}d_{124}\\d_{134}\end{pmatrix},\\
\mathbf g_{\mathrm m}&=\begin{pmatrix}\Psi_{124}-Q_{124}\\\Psi_{134}-Q_{134}\end{pmatrix},
\end{aligned}
\label{eq:3222-main-reduced-system}
\end{equation}
where $d_{124}=\mathcal D_{124}/(\alpha_2-\alpha_4)$ and $d_{134}=\mathcal D_{134}/(\alpha_3-\alpha_4)$. The short matrix reduction and the chosen cycle orientations are recorded in Appendix~\ref{app:3222-reduction}.

For pairwise distinct real parts, $d_{124}=d_{134}=0$ is equivalent to collinearity of the four auxiliary spectral points. Hence, if $(d_{124},d_{134})\ne(0,0)$, the auxiliary spectral points are noncollinear. In this case $\rho_{\Gamma_{\mathrm m}}=1$, and Eq.~\eqref{eq:3222-main-reduced-system} gives two scalar equations for the single variable $T$. They admit a common solution if and only if
\begin{equation}
d_{134}(\Psi_{124}-Q_{124})
-
d_{124}(\Psi_{134}-Q_{134})
=
0.
\label{eq:3222-explicit-realizability}
\end{equation}
When this selection condition holds, $T$ is uniquely determined, consistent with $D_{\mathrm{int}}=0$.

If instead $d_{124}=d_{134}=0$, the auxiliary spectral points are collinear. Equation~\eqref{eq:3222-main-reduced-system} then becomes independent of $T$, and realizability requires the two selection conditions
\begin{equation}
\Psi_{124}=Q_{124},
\qquad
\Psi_{134}=Q_{134}.
\label{eq:3222-degenerate-selection}
\end{equation}
When they hold, $T$ remains free, consistent with $D_{\mathrm{int}}=1$ and $\rho_{\Gamma_{\mathrm m}}=2$.

\paragraph{Fully pairwise collision.}
For $\Gamma_{\mathrm p}=\{12,13,14,23,24,34\}$, no additional explicit reduction is required. Applying Corollary~\ref{cor:pairwise-rigidity} at $N=4$ gives
\[
(D_{\mathrm{int}},\rho_{\Gamma_{\mathrm p}})
=
(0,1)
\]
for noncollinear spectral points and
\[
(D_{\mathrm{int}},\rho_{\Gamma_{\mathrm p}})
=
(2,3)
\]
on the collinear locus. Accordingly, the general spectral selection rule \eqref{eq:spectral-selection-rule} supplies one independent compatibility condition in the noncollinear case and three in the collinear case. When these conditions hold, the algebraically compatible relative geometry is unique in the former case and two-dimensional in the latter.

\begin{figure*}[t]
\centering
\includegraphics[width=\linewidth]
{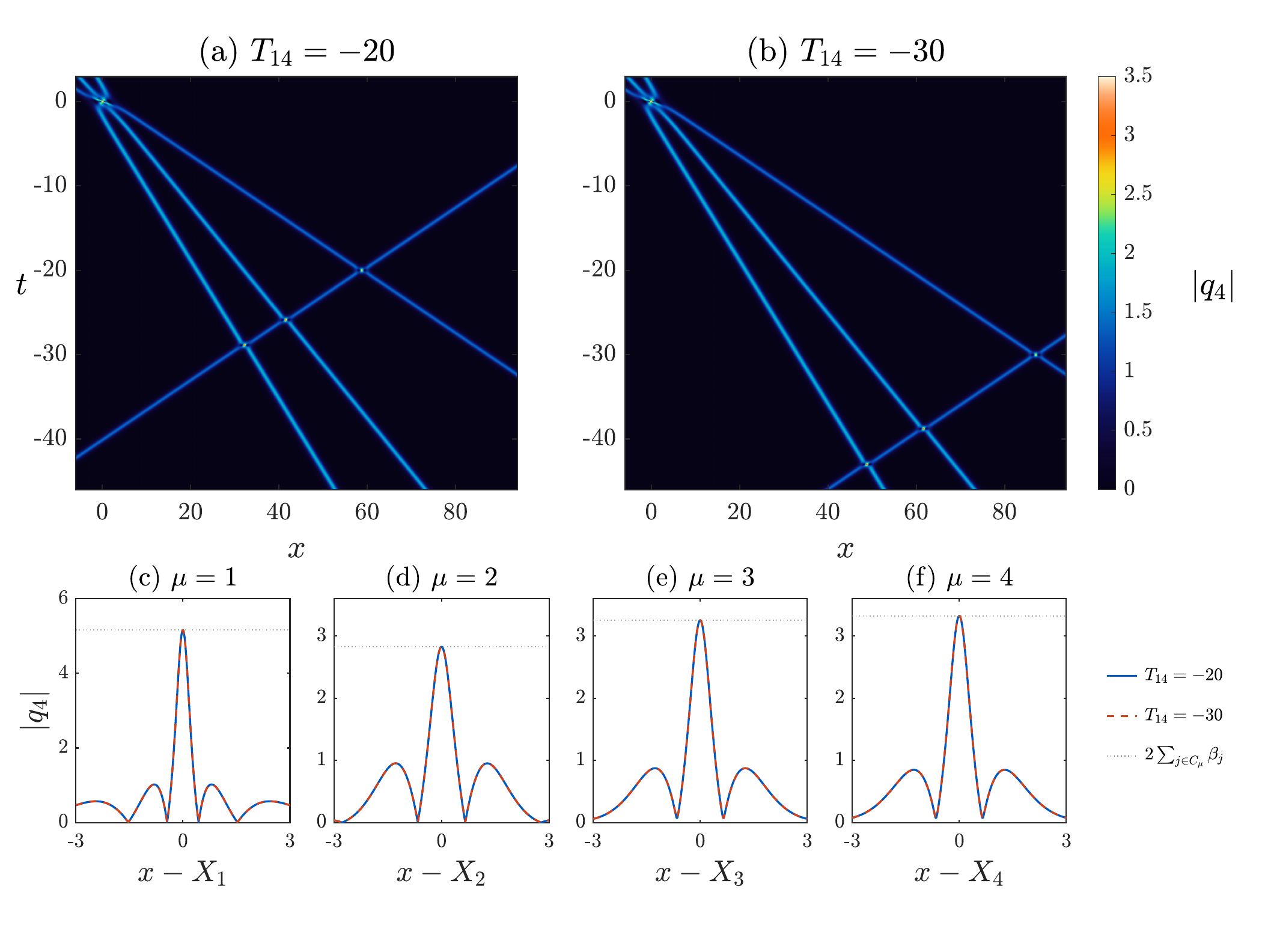} \caption{Fixed-spectrum internal deformation of the mixed four-soliton organization $\Gamma_{\mathrm m}=\{123,14,24,34\}$, with the unit-circle spectrum specified in Appendix~\ref{app:mixed-numerical-construction}. Panels (a) and (b) show $|q_4(x,t)|$ for $T_{14}=-20$ and $-30$, respectively. The sector assignment and integer branch $(n_{124},n_{134})=(-1,-2)$ are fixed. The family is parametrized by $T_{14}$. Panels (c)--(f) show retarded-time profiles at the propagation distances of events $123$, $14$, $24$, and $34$, respectively, plotted against $x-X_\mu$. Solid blue and dashed orange curves correspond to $T_{14}=-20$ and $-30$, respectively. Dotted gray lines mark $2\sum_{j\in\mathcal C_\mu}\beta_j$.}
\label{fig:mixed-flexible-field-sections}
\end{figure*}

\paragraph{Flexible mixed realization.}
For the unit-circle spectrum, sectors, and fixed phase branch specified in Appendix~\ref{app:mixed-numerical-construction}, the two selection conditions~\eqref{eq:3222-degenerate-selection} hold. Therefore $T_{14}$ remains a free internal deformation parameter. Figure~\ref{fig:mixed-flexible-field-sections} compares two members of this family with the same spectrum, sectors, and integer branch but different reconstructed norming data. The interval $T_{14}\in[-30,-20]$ satisfies the admissibility checks at isolation tolerance $\varepsilon=2\times10^{-6}$. With the constituent soliton parameters fixed, varying $T_{14}$ changes the relative event geometry while preserving the prescribed synchronization conditions.

\subsection{Constrained Fourier interference in the mixed flexible family}
\label{subsec:fourier-consequence}
\label{sec:local-structure-spectral-signatures}

The internal deformation of the mixed family has a Fourier signature fixed by the synchronization constraints. In the separated-cluster regime, these constraints link the cluster separation to the relative phase and thereby determine how the interference fringes vary along the family. We use the fixed spectrum and norming data reconstruction of Appendix~\ref{app:mixed-numerical-construction}, with $P_{123}=(0,0)$ and $\tau=T_{14}=-30,-25,-20$. All three samples lie in the certified interval $[-30,-20]$. The spectrum and reconstruction branch remain unchanged. The parameter $\tau$ labels different solutions, all evaluated at $t=0$.

\subsubsection{Effective cluster profiles and two-cluster approximation}

At $t=0$, the synchronized cluster $\mathcal C_{123}=\{1,2,3\}$ is centered at $x=0$, while soliton~$4$ lies to its right. We determine their effective profiles separately.

For $j\in\mathcal C_{123}$, substituting $\epsilon_4^{(123)}=-1$ and $P_{123}=(0,0)$ into Eq.~\eqref{eq:local-effective-state}, with $2g_j(0,0)=c_j$ from Eq.~\eqref{eq:half-exponent}, gives
\[
\mathcal Z_{j|123}^{\mathrm{eff}}(0,0)
=c_j-\Omega_{j4}.
\]
The synchronization conditions~\eqref{eq:local-sync} therefore require
\[
c_j-\Omega_{j4}
=\ii\Phi_{123}+2 n_j\pi\ii,
\qquad n_j\in\mathbb Z,\quad j=1,2,3.
\]
Choosing $\Phi_{123}=0$ and $n_j=0$ yields $c_j=\Omega_{j4}$. Equation~\eqref{eq:effective-local-g} then reduces to $\widetilde g_{j|123}(x,0)=-\ii\kappa_jx$. Thus, with $\kappa_1,\kappa_2,\kappa_3$ fixed, the effective three-soliton profile is the $\tau$-independent $q_{123}^{\mathrm{sync}}(x)$, as shown by Eq.~\eqref{eq:fourier-centered-cluster-identity} in Appendix~\ref{app:fourier-core-symmetry}.

For the singleton cluster $\{4\}$, the exterior sector signs are $\epsilon_j=+1$, $j=1,2,3$. Substituting $g_4$ from Eq.~\eqref{eq:half-exponent} into Eq.~\eqref{eq:effective-local-g} gives
\[
\widetilde g_4(x,t;\tau)
=
-\ii\kappa_4x-2\ii\kappa_4^2t
+\frac12\left[c_4(\tau)+\sum_{j=1}^{3}\Omega_{4j}\right].
\]
Here $c_4(\tau)=s_4(\tau)+\ii\vartheta_4(\tau)$ is the norming parameter defined in Eq.~\eqref{eq:norming-exponents}, reconstructed from the synchronization conditions at event $14$. We therefore define
\begin{equation}
H_4:=\sum_{j=1}^{3}\Omega_{4j},
\qquad
c_4^{\mathrm{eff}}(\tau):=c_4(\tau)+H_4.
\label{eq:fourier-effective-norming}
\end{equation}

Accordingly, the effective position and phase parameters are $\operatorname{Re}c_4^{\mathrm{eff}}(\tau)$ and $\operatorname{Im}c_4^{\mathrm{eff}}(\tau)$, respectively. Equation~\eqref{eq:bare-center} then gives the envelope center $x=D(\tau)$ of soliton~$4$ at $t=0$. Evaluating Eq.~\eqref{eq:single-carrier-phase} at $(x,t)=(D(\tau),0)$ gives the carrier phase $\phi(\tau)$ at that center:
\begin{equation}
\begin{aligned}
D(\tau)&:=-\frac{\operatorname{Re}c_4^{\mathrm{eff}}(\tau)}{2\beta_4},\\
\phi(\tau)&:=\operatorname{Im}c_4^{\mathrm{eff}}(\tau)-2\alpha_4D(\tau)\pmod{2\pi}.
\end{aligned}
\label{eq:fourier-distance-phase}
\end{equation}
For the selected family, the reconstruction in Appendix~\ref{app:fourier-phase-rates} gives $D(\tau)>0$ throughout $\tau\in[-30,-20]$, so soliton~$4$ lies to the right of the $123$ core. Since the $123$ core is centered at $x=0$ with $\Phi_{123}=0$, $D(\tau)$ is the distance between the two cluster centers, and $\phi(\tau)$ is their phase difference evaluated at their respective centers. In the optical interpretation, $D(\tau)$ is the retarded-time separation of the two clusters at the fixed propagation distance $t=0$.

Using Eq.~\eqref{eq:fourier-distance-phase}, we rewrite the effective single-soliton field from Eq.~\eqref{eq:single-soliton} at $t=0$ as $e^{\ii\phi(\tau)}Q_4(x-D(\tau))$, where
\[
Q_4(\xi)
=
2\beta_4\sech(2\beta_4\xi)e^{-2\ii\alpha_4\xi}
\]
is its centered profile. Since $\kappa_4$ is fixed, $Q_4$ is independent of $\tau$.

Combining the two effective cluster profiles, we obtain the following approximation when the clusters are well separated:
\begin{equation}
q_4(x,0;\tau)
\approx
q_{123}^{\mathrm{sync}}(x)
+
e^{\ii\phi(\tau)}Q_4(x-D(\tau)).
\label{eq:fourier-two-cluster-field}
\end{equation}
The local reduction~\eqref{eq:local-reduction} motivates this approximation near each cluster.

\subsubsection{Separation--phase constraint and Fourier interference}

To determine how the two-cluster approximation varies with $\tau$, we use the synchronization conditions to calculate $D'(\tau)$ and $\phi'(\tau)$. With a continuous representative of $\phi(\tau)$, the derivation in Appendix~\ref{app:fourier-phase-rates} gives
\begin{equation}
D'=v=4(\alpha_4-\alpha_1),\qquad
\phi'=\omega=4(|\kappa_1|^2-|\kappa_4|^2),
\label{eq:fourier-exact-rates}
\end{equation}
where primes denote derivatives with respect to $\tau$. For collinear auxiliary spectral points, write $|\kappa_j|^2=a_{\mathrm{sp}}\alpha_j+b$, where $a_{\mathrm{sp}}$ and $b$ are constants. Equation~\eqref{eq:fourier-exact-rates} then gives $\phi'=-a_{\mathrm{sp}}D'$. Integrating with respect to $\tau$ gives
\begin{equation}
\phi(\tau)+a_{\mathrm{sp}}D(\tau)=\chi,
\label{eq:fourier-collinear-phase-lock}
\end{equation}
where $\chi$ is constant along the family. Thus, $D(\tau)$ and $\phi(\tau)$ cannot be varied independently while preserving the prescribed synchronization conditions.

For the unit-circle spectrum~\eqref{eq:flexible-circle-family}, $a_{\mathrm{sp}}=0$. With $\theta_1=\pi/4$ and $\theta_4=3\pi/4$, Eq.~\eqref{eq:fourier-exact-rates} gives $D'=-4\sqrt{2}$ and $\phi'=0$. Thus, the separation decreases as $\tau$ increases, while the centered relative phase remains fixed.

We use the Fourier transform
\[
\widehat f(k):=\int_{\mathbb R}f(x)e^{-\ii kx}\,\dd x.
\]
Here $k$ is the normalized angular frequency offset conjugate to the retarded time $x$. We denote the transforms of the centered profiles by
\[
\mathcal F_{123}(k)
:=\widehat{q_{123}^{\mathrm{sync}}}(k),
\qquad
\mathcal F_4(k):=\widehat Q_4(k).
\]
Both $\mathcal F_{123}(k)$ and $\mathcal F_4(k)$ are independent of $\tau$.

The transform $\mathcal F_{123}(k)$ is evaluated from the exact synchronized three-soliton profile. As shown in Appendix~\ref{app:fourier-core-symmetry}, the symmetry $q_{123}^{\mathrm{sync}}(-x) =\overline{q_{123}^{\mathrm{sync}}(x)}$ implies that $\mathcal F_{123}(k)$ is real. Direct transformation of $Q_4$ gives
\begin{equation}
\mathcal F_4(k)
=\pi\sech\!\left[
\frac{\pi(k+2\alpha_4)}{4\beta_4}
\right]>0.
\label{eq:fourier-singleton-transform}
\end{equation}

Let $q_{\mathrm{int}}(x;\tau)$ denote the two-cluster approximation given by the right-hand side of Eq.~\eqref{eq:fourier-two-cluster-field}. Applying the Fourier translation identity gives
\[
\widehat q_{\mathrm{int}}(k;\tau)
=\mathcal F_{123}(k)
+e^{\ii\phi(\tau)-\ii kD(\tau)}\mathcal F_4(k).
\]
We denote the Fourier power spectra of the full four-soliton field and the two-cluster approximation by
\[
P(k;\tau):=|\widehat q_4(k,0;\tau)|^2,
\qquad
P_{\mathrm{int}}(k;\tau)
:=|\widehat q_{\mathrm{int}}(k;\tau)|^2,
\]
respectively.

Expanding $P_{\mathrm{int}}$ gives
\begin{equation}
\begin{aligned}
P_{\mathrm{int}}(k;\tau)
={}&\mathcal F_{123}(k)^2+\mathcal F_4(k)^2\\
&+2\mathcal F_{123}(k)\mathcal F_4(k)\cos\Psi(k;\tau).
\end{aligned}
\label{eq:fourier-constrained-interference}
\end{equation}
Here $\Psi(k;\tau)$ denotes the Fourier interference phase. Using the synchronization constraint \eqref{eq:fourier-collinear-phase-lock}, we have
\begin{equation}
\Psi(k;\tau)
:=kD(\tau)-\phi(\tau)
=(k+a_{\mathrm{sp}})D(\tau)-\chi.
\label{eq:fourier-interference-phase}
\end{equation}
Differentiating Eq.~\eqref{eq:fourier-interference-phase} with respect to $\tau$ gives
\[
\partial_\tau\Psi=(k+a_{\mathrm{sp}})D',
\]
which describes how the interference phase varies along the family.

At fixed $\tau$, let $\Delta k(\tau)>0$ denote the period in normalized angular frequency $k$ of the cosine interference factor. Then
\[
\Psi(k+\Delta k(\tau);\tau)-\Psi(k;\tau)
=D(\tau)\Delta k(\tau)=2\pi,
\]
which yields
\[
\Delta k(\tau)=\frac{2\pi}{D(\tau)}.
\]
For the unit-circle realization, this period increases as the cluster separation decreases, while the discrete eigenvalues remain fixed.

\subsubsection{Comparison with the full four-soliton field}

\begin{figure*}[t]
\centering
\includegraphics[width=0.78\linewidth]{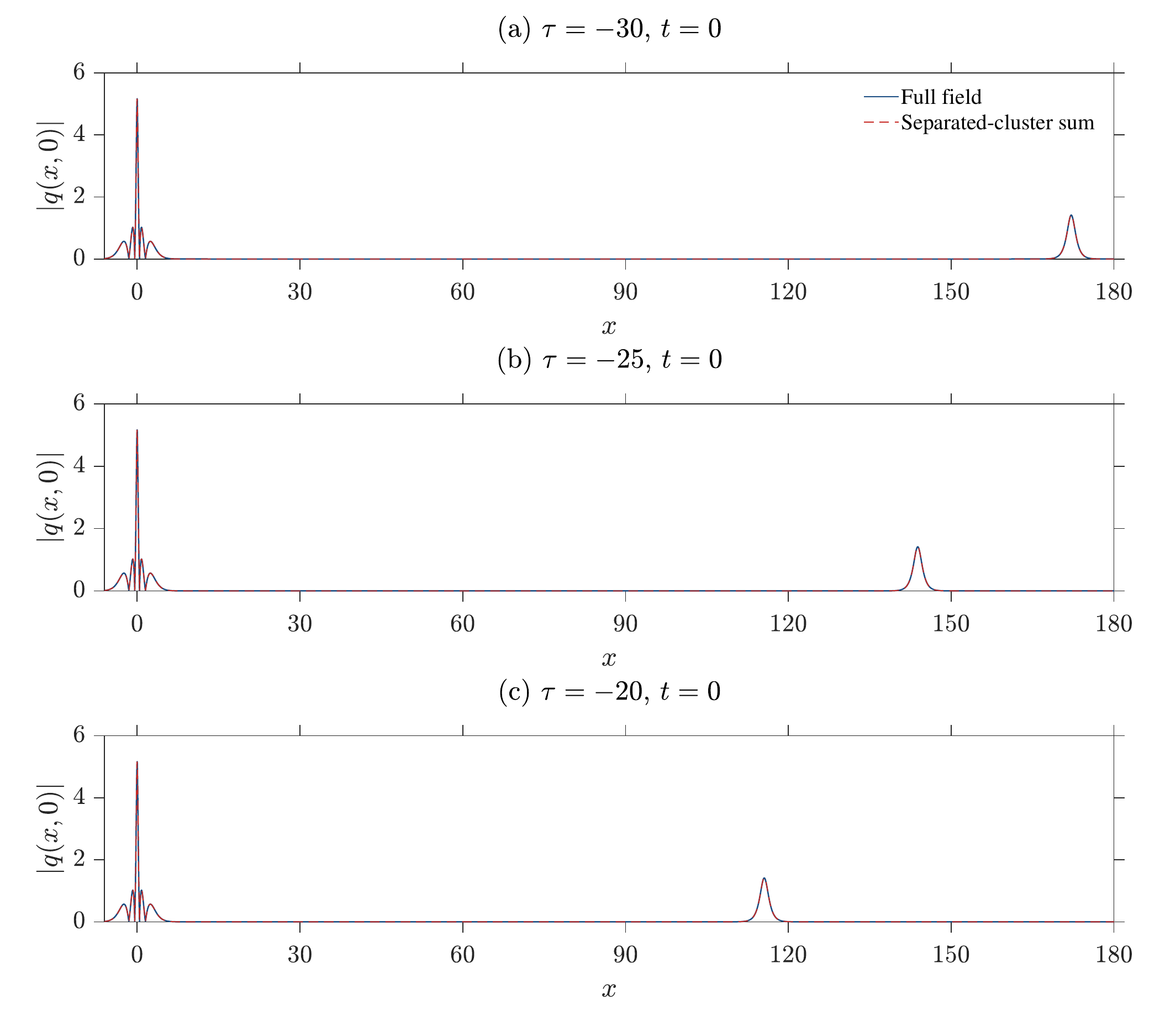}
\caption{Retarded-time profiles at $t=0$ for (a) $\tau=-30$, (b) $\tau=-25$, and (c) $\tau=-20$, using the fixed spectrum~\eqref{eq:flexible-circle-family}. Solid blue curves show the full four-soliton amplitude $|q_4(x,0;\tau)|$, and dashed red curves give the modulus of the two-cluster approximation \eqref{eq:fourier-two-cluster-field}. The common retarded-time scale highlights the changing position of soliton~$4$ relative to the synchronized cluster of solitons 1--3.}
\label{fig:fourier-spatial-sections}
\end{figure*}

We assess the two-cluster approximation by comparing its retarded-time profiles and Fourier power spectra with those computed from the full four-soliton field. Figure~\ref{fig:fourier-spatial-sections} compares the retarded-time profiles at $t=0$. The reconstructed separations are $D\approx172.131,143.847,115.563$ for $\tau=-30,-25,-20$, respectively.

\begin{figure*}[t]
\centering
\includegraphics[width=0.75\linewidth]{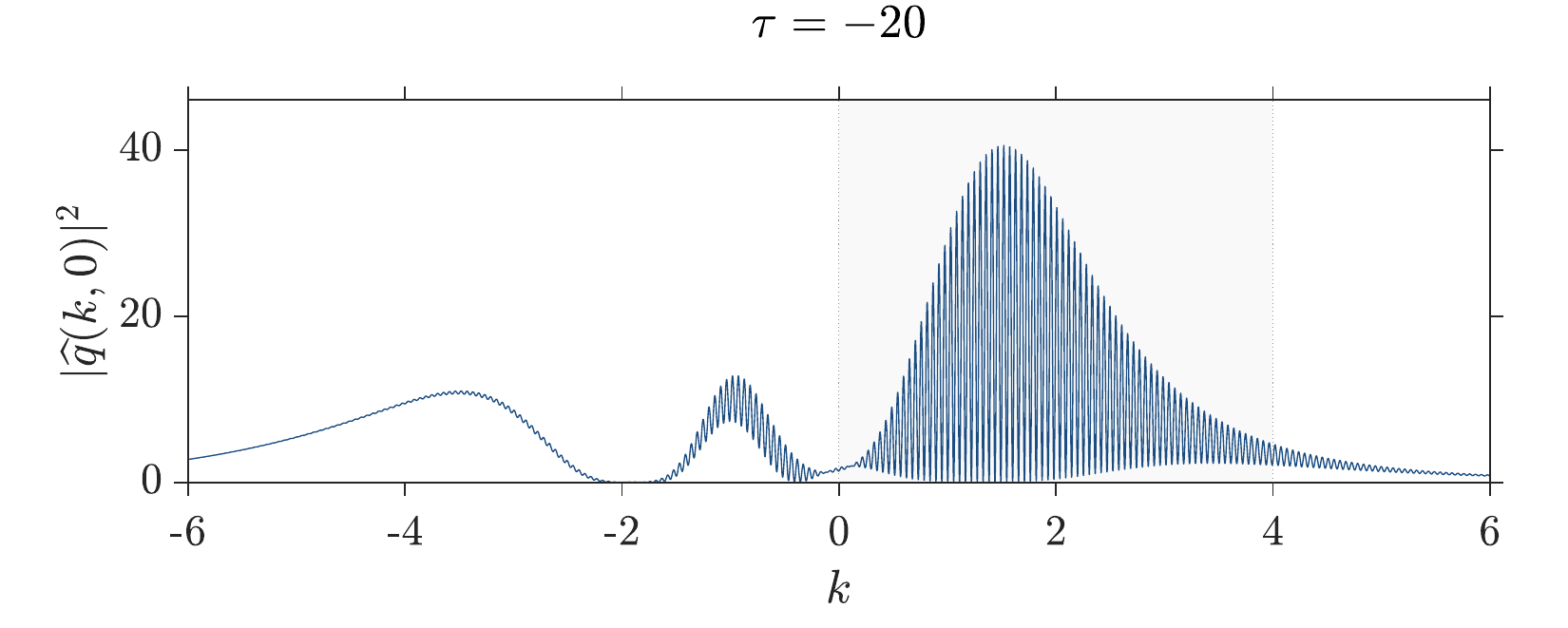}
\caption{Fourier power overview at $t=0$ for $\tau=-20$ on $k\in[-6,6]$. The blue curve is computed from the full four-soliton field. The shaded interval $0\le k\le4$ is displayed at larger scale in Fig.~\ref{fig:fourier-flexible-family}(c).}
\label{fig:fourier-overview}
\end{figure*}

\begin{figure*}[t]
\centering
\includegraphics[width=0.88\linewidth]{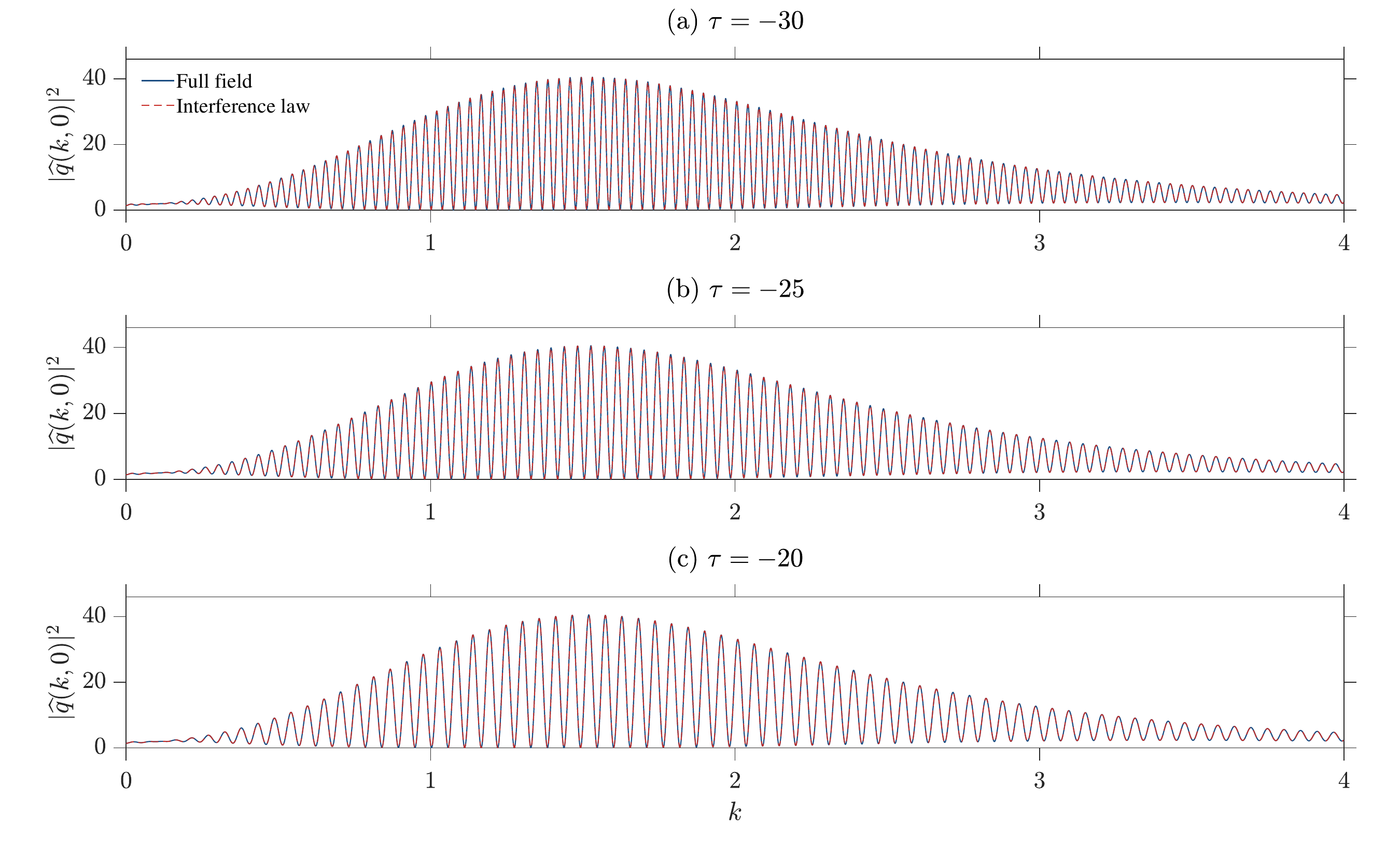}
\caption{Fourier interference at $t=0$ for (a) $\tau=-30$, (b) $\tau=-25$, and (c) $\tau=-20$. Solid blue curves show the full four-soliton Fourier power, and dashed red curves give the prediction \eqref{eq:fourier-constrained-interference}. The displayed normalized angular frequency range is $k\in[0,4]$. The discrete spectrum is fixed. The interference-factor periods $2\pi/D$ are approximately $0.0365$, $0.0437$, and $0.0544$, respectively.}
\label{fig:fourier-flexible-family}
\end{figure*}

Figure~\ref{fig:fourier-overview} shows the broader Fourier power spectrum for $\tau=-20$. Figure~\ref{fig:fourier-flexible-family} compares the predicted Fourier power spectra with those obtained from direct Fourier transforms of the full four-soliton fields for $\tau=-30,-25,-20$. Both $D$ and $\phi$ are obtained from the reconstructed norming data through Eq.~\eqref{eq:fourier-distance-phase}.

\begin{figure*}[t]
\centering
\includegraphics[width=0.42\linewidth,trim=0 12 0 28,clip]{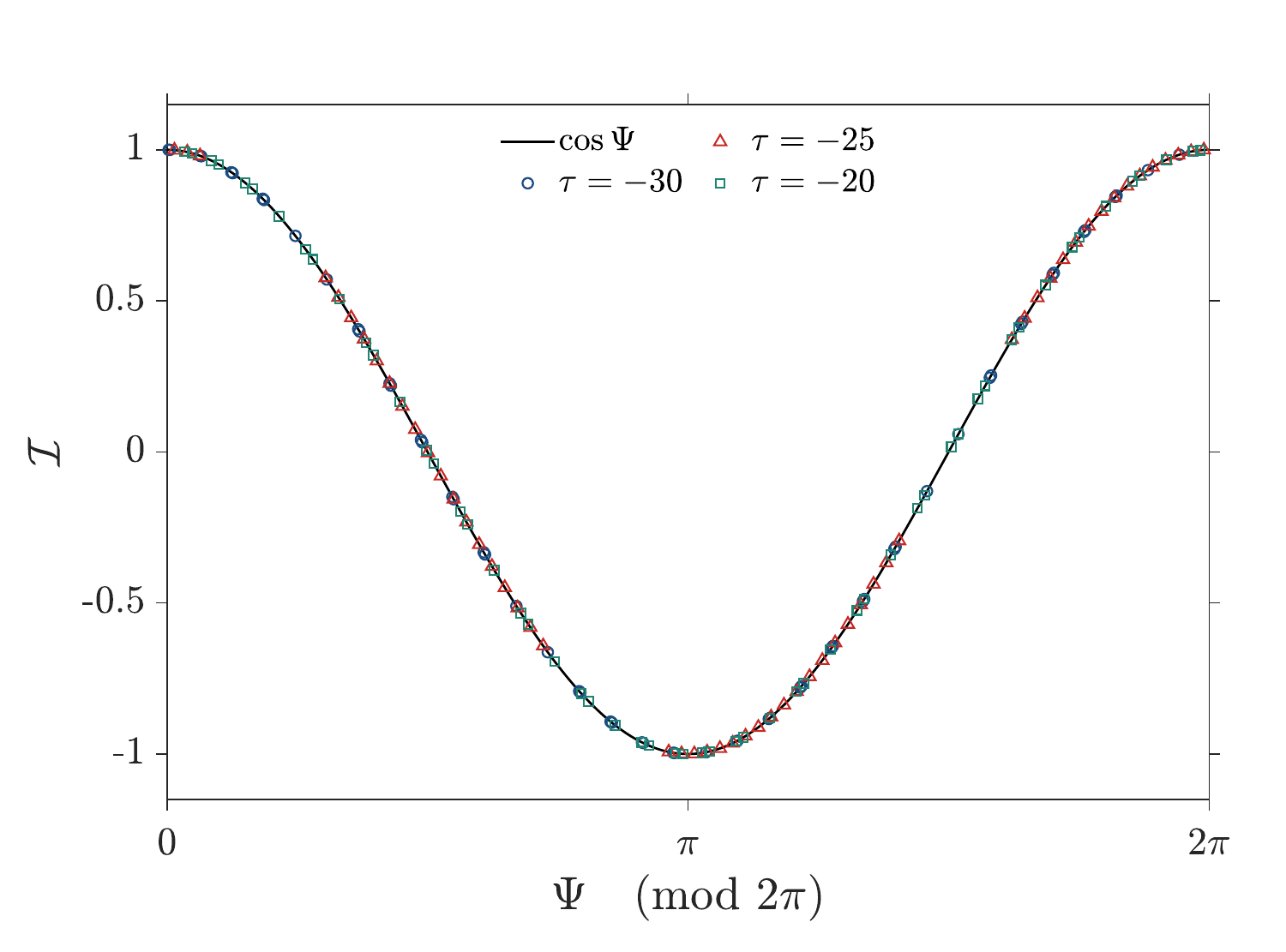}
\caption{Normalized full-field interference \eqref{eq:fourier-normalized-interference} on $I_k=[1.6,2.4]$, plotted against $\Psi\pmod{2\pi}$ for $\tau=-30,-25,-20$. Symbols are computed from the full four-soliton fields, and the solid black curve shows $\cos\Psi$.}
\label{fig:fourier-phase-comparison}
\end{figure*}

We denote the discrepancy between the full and predicted power spectra by $E(k;\tau):=P(k;\tau)-P_{\mathrm{int}}(k;\tau)$. To isolate the interference factor in Eq.~\eqref{eq:fourier-constrained-interference}, we define the normalized full-field interference $\mathcal I(k;\tau)$ on $I_k=[1.6,2.4]$ by
\begin{equation}
\begin{aligned}
\mathcal I(k;\tau)
&:=\frac{P(k;\tau)-\mathcal F_{123}(k)^2-\mathcal F_4(k)^2}
{2\mathcal F_{123}(k)\mathcal F_4(k)}\\
&=\cos\Psi(k;\tau)+\frac{E(k;\tau)}{2\mathcal F_{123}(k)\mathcal F_4(k)}.
\end{aligned}
\label{eq:fourier-normalized-interference}
\end{equation}
The interval $I_k=[1.6,2.4]$ is chosen to avoid small values of the denominator $2\mathcal F_{123}(k)\mathcal F_4(k)$. Figure~\ref{fig:fourier-phase-comparison} plots $\mathcal I$ against $\Psi\pmod{2\pi}$ to test the predicted phase dependence across the three family members. The normalized interference agrees closely with the predicted cosine dependence for all three samples. The numerical discrepancies and convergence checks are reported in Appendix~\ref{app:fourier-numerical-verification}.

\section{Discussion}
\label{sec:discussion}

The present results clarify two central aspects of synchronized collision patterns: the spectral constraints on the joint realizability of multiple prescribed collisions within a single multisoliton solution and the freedom to vary their relative arrangement in retarded time and propagation distance at fixed eigenvalues while preserving synchronization.

For distinct discrete eigenvalues, we derive an affine system of compatibility equations for the collision event positions by requiring all events to share the same norming data. Given a discrete spectrum and collision organization, the solvability of this system determines whether a common set of norming data can satisfy the synchronization conditions at all events and thus whether the spectrum supports the organization algebraically. Since the discrete eigenvalues determine the soliton amplitudes and velocities, these properties constrain which collision organizations can be synchronized by adjusting the soliton positions and phases. The physical admissibility conditions further identify which compatible configurations correspond to sufficiently isolated collision events.

Even when a collision organization is realizable, the synchronization conditions may fix the relative positions of its collision events. For complete collision organizations, we use the discrete spectrum to characterize how the synchronization conditions constrain continuous changes in the relative event positions. When the auxiliary points $(|\kappa_j|^2,\operatorname{Re}\kappa_j)$ are noncollinear, maintaining synchronization during a continuous deformation with the discrete eigenvalues and complete collision organization fixed allows all events to shift together, but leaves their relative separations in retarded time and propagation distance unchanged. The three-soliton example illustrates that such rigidity does not imply a unique collision organization: at the same discrete spectrum, different norming data realize either a collective three-soliton event or three resolved pairwise collision events.

When the auxiliary spectral points are collinear, the available freedom depends on how the solitons are grouped into collision events. For the two classes considered here with $N\geq3$, explicit internal deformation dimensions have been obtained: fully pairwise organizations have $N-2$ internal algebraic degrees of freedom, whereas organizations with one $m$-soliton collective event and all remaining interactions pairwise have $N-m$, with $3\leq m\leq N$. These dimensions quantify the tunability of the relative event positions at fixed discrete eigenvalues, specifying the number of independent parameters that can be continuously varied while preserving synchronization.

The mixed four-soliton family illustrates how synchronized collision events can change their relative spacetime positions at fixed discrete eigenvalues while retaining their leading-order center amplitudes. This deformation is also reflected in the Fourier power spectrum. For this family, synchronization and the collinearity of the auxiliary spectral points impose an affine relation between cluster separation and relative phase, constraining the interference phase in the two-cluster approximation. In the unit-circle realization, the relative phase at the cluster centers remains fixed as their separation $D$ at $t=0$ varies, changing the period $2\pi/D$ of the oscillatory interference factor. Thus, fixed discrete eigenvalues do not imply a fixed ordinary Fourier interference pattern.

More generally, the local reduction applies to sufficiently isolated synchronized events in an $N$-soliton field. Its leading-order term is an effective field of the participating solitons that incorporates the finite position and phase shifts induced by remote solitons, while the remainder is exponentially small in the isolation parameter. Under the synchronization and isolation conditions, the field amplitude at each prescribed center therefore equals the sum of the participating soliton amplitudes up to exponentially small corrections. At fixed discrete eigenvalues and participating sets, this amplitude sum remains unchanged along admissible internal deformations. The exponentially small corrections depend on the isolation of each event.

The local reduction also provides a way to compare the profiles near isolated collisions with those of rational rogue waves. Intriguingly, selected two-, three-, and four-soliton collisions have been shown to exhibit local amplitude profiles at the time of their maximum elevation that resemble first-, second-, and third-order Peregrine solitons, respectively~\cite{Gelash2018}. These examples motivate a formal comparison between an $N$-soliton collision and an $n$th-order rogue wave with $n=N-1$. Under this correspondence, a complete pairwise organization of $N$ solitons contains $N(N-1)/2$ collision events, equal to the $n(n+1)/2$ fundamental peaks obtained when an $n$th-order rogue wave is fully resolved~\cite{YangYang2021}. Each sufficiently isolated synchronized pairwise event has an effective two-soliton field. When the effective parameters yield a first-order Peregrine-like amplitude profile, the local reduction formula transfers this approximation to the full multisoliton field within their common region of validity, with an additional exponentially small error. The comparison therefore concerns both the number of resolved events and the local amplitude profiles they can exhibit. In the multisoliton setting, all events must share common scattering data, so their joint realization is subject to the compatibility and admissibility conditions established here.

Within the integrable NLS description of optical pulse propagation, these findings provide a theoretical foundation for engineering multisoliton collision patterns in optical fibers, clarifying which prescribed configurations can be realized and under what conditions their relative spacetime geometry can be continuously adjusted while preserving synchronization. Moreover, the connection between geometric adjustments and Fourier interference demonstrated in the four-soliton family identifies spectral signatures that could be used to characterize such changes in optical-fiber experiments.

A natural extension of this framework is to investigate the realizability of synchronized collision organizations and the tunability of their relative event positions in the large-$N$ limit. Another direction is to extend the local analysis to events that are not well separated and quantify how overlap between neighboring collision regions affects the local wave profiles and amplitude enhancement.

\begin{acknowledgments}
We are grateful to each member of our discussion group for their suggestions. This work was supported by the National Natural Science Foundation of China under Grant No.12575005, the National Natural Science Foundation of China under Grant No. 12472387, and the Shanxi Province Science Foundation under Grant No. 202303021221031.
\end{acknowledgments}

\appendix

\section{Local reduction of an isolated synchronized collision}
\label{app:local_reduction}

We establish the local reduction~\eqref{eq:local-reduction} and the amplitude-sum law~\eqref{eq:local-amplitude-sum} for an isolated event satisfying the synchronization conditions~\eqref{eq:local-sync}. Throughout this appendix, the eigenvalues $\kappa_j$ are fixed and distinct, with $\operatorname{Im}\kappa_j>0$. All error estimates are uniform on the fixed neighborhood $\mathcal U_\mu$ defined in Eq.~\eqref{eq:local-neighborhood}. If $\mathcal R_\mu=\varnothing$, the reduction is exact. Empty matrix blocks are omitted below.

\paragraph{Elimination of the remote block.}
For the calculations in this appendix, introduce
\[
K_{jm}=\frac{1}{\kappa_j^*-\kappa_m},\qquad
\mathcal R_{\mu,\pm}
=\{r\in\mathcal R_\mu:\epsilon_r^{(\mu)}=\pm1\},
\]
and let $\bm K_{S,S}$ denote the submatrix indexed by $S$. The sector signs retain their definition in Eq.~\eqref{eq:event-sector-sign}.

For each $r\in\mathcal R_\mu$, define $ C_r:=\sum_{\ell\ne r} \bigl|\operatorname{Re}\Omega_{r\ell}\bigr|. $ By Eqs.~\eqref{eq:remote-local-coordinate} and~\eqref{eq:isolation-measure}, the triangle inequality gives
\[
2L_\mu
\le
\left|
\operatorname{Re}\mathcal Z_{r|\mu}^{\mathrm{loc}}(P_\mu)
\right|
\le
2\left|\operatorname{Re}g_r(P_\mu)\right|+C_r.
\]
Hence
\begin{equation}
\left|\operatorname{Re}g_r(P_\mu)\right|
\ge L_\mu-\frac{C_r}{2},
\qquad r\in\mathcal R_\mu.
\label{eq:isolation-to-bare-exponent}
\end{equation}
For a fixed spectrum, $C_r$ is constant, so increasing $L_\mu$ forces every remote bare exponent away from zero.

The affine form~\eqref{eq:half-exponent} implies
\[
|\operatorname{Re}g_r(x,t)-\operatorname{Re}g_r(P_\mu)|
\le\beta_r h_x+4|\alpha_r|\beta_r h_t.
\]
Together with Eq.~\eqref{eq:isolation-to-bare-exponent}, this yields $2\epsilon_r^{(\mu)}\operatorname{Re}g_r(x,t) \ge 2L_\mu-C_r-2\beta_rh_x-8|\alpha_r|\beta_rh_t$. Hence
\begin{equation}
 z_r:=e^{-2\epsilon_r^{(\mu)}g_r}=O(e^{-2L_\mu}),
 \qquad r\in\mathcal R_\mu,
 \label{app:remote-small-branch}
\end{equation}
on $\mathcal U_\mu$. The remote signs are fixed there for sufficiently large $L_\mu$. The synchronization conditions~\eqref{eq:local-sync} and Eq.~\eqref{eq:effective-local-g} also imply that $\operatorname{Re}g_j$, $j\in\mathcal C_\mu$, remains uniformly bounded on this neighborhood.

In both matrices~\eqref{eq:G-matrix} and~\eqref{eq:H-matrix}, multiply the row and column corresponding to each remote index $r$ by $e^{-\epsilon_r^{(\mu)}g_r^*}$ and $e^{-\epsilon_r^{(\mu)}g_r}$, respectively. Denote the resulting matrices by $\widehat{\bm G}$ and $\widehat{\bm H}$. The common factors cancel from the determinant ratio in Eq.~\eqref{eq:N-soliton-determinant}. Direct substitution gives
\begin{equation}
\widehat G_{rs}=
\begin{cases}
K_{rs}(1+z_r^*z_s),&\epsilon_r^{(\mu)}=\epsilon_s^{(\mu)},\\
K_{rs}(z_r^*+z_s),&\epsilon_r^{(\mu)}\ne\epsilon_s^{(\mu)},
\end{cases}
\label{app:normalized-remote-entries}
\end{equation}
and, for $j,m\in\mathcal C_\mu$,
\begin{equation}
\begin{aligned}
\widehat G_{jr}
 &=K_{jr}e^{\epsilon_r^{(\mu)}g_j^*}+O(e^{-2L_\mu}),\\
\widehat G_{rm}
 &=K_{rm}e^{\epsilon_r^{(\mu)}g_m}+O(e^{-2L_\mu}).
\end{aligned}
\label{app:normalized-cross-entries}
\end{equation}
Thus the remote block has the form
\begin{equation}
\begin{aligned}
\bm D&:=\widehat{\bm G}_{\mathcal R_\mu,\mathcal R_\mu}
      =\bm D_0+O(e^{-2L_\mu}),\\
\bm D_0&=
\begin{pmatrix}
\bm K_{\mathcal R_{\mu,+},\mathcal R_{\mu,+}}&0\\
0&\bm K_{\mathcal R_{\mu,-},\mathcal R_{\mu,-}}
\end{pmatrix}.
\end{aligned}
\label{app:remote-block-leading}
\end{equation}
Each nonempty diagonal block is nonsingular by the Cauchy determinant formula. Since $\bm D_0$ is fixed, a Neumann expansion yields
\begin{equation}
\bm D^{-1}=\bm D_0^{-1}+O(e^{-2L_\mu})
\label{Dni}
\end{equation}
for sufficiently large $L_\mu$.

The Cauchy determinant formula yields the following identities:
\begin{align}
K_{jm}-\bm K_{j,S}\bm K_{S,S}^{-1}\bm K_{S,m}
 &=K_{jm}e^{\sum_{r\in S}(\Omega_{jr}^*+\Omega_{mr})},
 \label{app:cauchy-schur}\\
1-\bm1^{\mathsf T}\bm K_{S,S}^{-1}\bm K_{S,m}
 &=e^{\sum_{r\in S}\Omega_{mr}},
 \label{app:cauchy-upper}\\
1-\bm K_{j,S}\bm K_{S,S}^{-1}\bm1
 &=e^{\sum_{r\in S}\Omega_{jr}^*},
 \label{app:cauchy-lower}
\end{align}
where $j,m\notin S$ and $\bm1$ is the all-ones column vector of the required dimension. Indeed, the left-hand side of Eq.~\eqref{app:cauchy-schur} equals
\[
\frac{\det\begin{pmatrix}
\bm K_{S,S}&\bm K_{S,m}\\\bm K_{j,S}&K_{jm}
\end{pmatrix}}{\det\bm K_{S,S}}
=K_{jm}\prod_{r\in S}
\frac{(\kappa_j^*-\kappa_r^*)(\kappa_m-\kappa_r)}
     {(\kappa_j^*-\kappa_r)(\kappa_m-\kappa_r^*)}.
\]
Equation~\eqref{eq:scattering-log-principal} identifies each factor with $e^{\Omega_{jr}^*+\Omega_{mr}}$. Treating the row variable $\kappa_j^*$ as an independent complex variable, multiplying by it, and taking it to infinity gives Eq.~\eqref{app:cauchy-upper}. The corresponding limit in the column variable, after multiplication by $-\kappa_m$, gives Eq.~\eqref{app:cauchy-lower}.

Order the soliton indices as $\mathcal C_\mu,\mathcal R_\mu$ and introduce the block notation
\[
\widehat{\bm G}=
\begin{pmatrix}
\bm A&\bm B\\
\bm C&\bm D
\end{pmatrix},
\qquad
\widehat{\bm H}=
\begin{pmatrix}
0&\bm a_{\mathcal C_\mu}^{\mathsf T}
 &\bm a_{\mathcal R_\mu}^{\mathsf T}\\
\bm b_{\mathcal C_\mu}&\bm A&\bm B\\
\bm b_{\mathcal R_\mu}&\bm C&\bm D
\end{pmatrix}.
\]
Here the vectors are the normalized border entries inherited from Eq.~\eqref{eq:H-matrix}. The entries of $\bm a_{\mathcal R_\mu}$ are $1$ on $\mathcal R_{\mu,+}$ and $z_r$ on $\mathcal R_{\mu,-}$, whereas those of $\bm b_{\mathcal R_\mu}$ are $z_r^*$ on $\mathcal R_{\mu,+}$ and $1$ on $\mathcal R_{\mu,-}$.

The Schur complements of $\bm D$ in $\widehat{\bm G}$ and $\widehat{\bm H}$ are $\bm S$ and $\left(\begin{smallmatrix} h&\widehat{\bm a}^{\mathsf T}\\ \widehat{\bm b}&\bm S
\end{smallmatrix}\right)$,
respectively, with
\begin{equation}
\begin{aligned}
\bm S&=\bm A-\bm B\bm D^{-1}\bm C,\\
h&=-\bm a_{\mathcal R_\mu}^{\mathsf T}\bm D^{-1}\bm b_{\mathcal R_\mu},\\
\widehat{\bm a}^{\mathsf T}&=\bm a_{\mathcal C_\mu}^{\mathsf T}
-\bm a_{\mathcal R_\mu}^{\mathsf T}\bm D^{-1}\bm C,\\
\widehat{\bm b}&=\bm b_{\mathcal C_\mu}
-\bm B\bm D^{-1}\bm b_{\mathcal R_\mu}.
\end{aligned}
\label{app:schur-complement-definitions}
\end{equation}
Here $\widehat{\bm a}=(\widehat a_m)_{m\in\mathcal C_\mu}$ and $\widehat{\bm b}=(\widehat b_j)_{j\in\mathcal C_\mu}$ denote the border vectors after elimination of the remote block. The hats on these vectors refer to this reduction. Using Eq.~\eqref{Dni} in the Schur complement $\bm S$ and applying Eq.~\eqref{app:cauchy-schur} to the two remote sectors gives
\begin{equation}
\begin{aligned}
S_{jm}={}&K_{jm}\Bigl[
e^{-g_j^*-g_m+\sum_{r\in\mathcal R_{\mu,-}}
(\Omega_{jr}^*+\Omega_{mr})}\\
&{}+e^{g_j^*+g_m+\sum_{r\in\mathcal R_{\mu,+}}
(\Omega_{jr}^*+\Omega_{mr})}\Bigr]+O(e^{-2L_\mu}).
\end{aligned}
\label{app:reduced-denominator-entries}
\end{equation}
Similarly, Eqs.~\eqref{app:cauchy-upper} and \eqref{app:cauchy-lower} yield
\begin{equation}
\begin{aligned}
\widehat a_m
&=e^{g_m+\sum_{r\in\mathcal R_{\mu,+}}\Omega_{mr}}
+O(e^{-2L_\mu}),\\
\widehat b_j
&=e^{-g_j^*+\sum_{r\in\mathcal R_{\mu,-}}\Omega_{jr}^*}
+O(e^{-2L_\mu}).
\end{aligned}
\label{app:reduced-border-entries}
\end{equation}
The leading contribution to $h$ vanishes because $\bm a_{\mathcal R_\mu}$ and $\bm b_{\mathcal R_\mu}$ have nonzero leading entries in different sectors, while $\bm D_0^{-1}$ is block diagonal. Hence $h=O(e^{-2L_\mu})$. The common factor $\det\bm D$ cancels, yielding the exact identity
\begin{equation}
q_N=-2\ii\,
\frac{\det\begin{pmatrix}
h&\widehat{\bm a}^{\mathsf T}\\
\widehat{\bm b}&\bm S
\end{pmatrix}}
{\det\bm S}.
\label{app:reduced-determinant-ratio}
\end{equation}

Set $ f_{j|\mu} =\exp\!\left( \frac12\sum_{r\in\mathcal R_\mu}\Omega_{jr} \right),$ $j\in\mathcal C_\mu$. The effective exponents defined in Eq.~\eqref{eq:effective-local-g} satisfy
\[
f_{j|\mu}e^{\pm\widetilde g_{j|\mu}}
=
e^{\pm g_j+
\sum_{r\in\mathcal R_{\mu,\pm}}\Omega_{jr}}.
\]
Let $\widetilde{\bm G}$ and $\widetilde{\bm H}$ denote the matrices obtained from Eqs.~\eqref{eq:G-matrix} and~\eqref{eq:H-matrix}, respectively, by restricting the soliton indices to $\mathcal C_\mu$ and replacing $g_j$ with $\widetilde g_{j|\mu}$. Using these identities in Eqs.~\eqref{app:reduced-denominator-entries} and~\eqref{app:reduced-border-entries} gives
\begin{equation}
\begin{aligned}
S_{jm}
&=f_{j|\mu}^*\widetilde G_{jm}f_{m|\mu}
  +O(e^{-2L_\mu}),\\
\widehat a_m
&=e^{\widetilde g_{m|\mu}}f_{m|\mu}
  +O(e^{-2L_\mu}),\\
\widehat b_j
&=f_{j|\mu}^*e^{-\widetilde g_{j|\mu}^*}
  +O(e^{-2L_\mu}).
\end{aligned}
\label{app:effective-factorization}
\end{equation}

Since $h=O(e^{-2L_\mu})$ and the factors $f_{j|\mu}$ and their reciprocals are bounded for fixed distinct eigenvalues, Eqs.~\eqref{app:reduced-determinant-ratio} and~\eqref{app:effective-factorization} yield
\begin{equation}
\begin{aligned}
q_N
&=-2\ii
\frac{
\left(\prod_{j\in\mathcal C_\mu}|f_{j|\mu}|^2\right)
\det(\widetilde{\bm H}+\bm E_H)
}{
\left(\prod_{j\in\mathcal C_\mu}|f_{j|\mu}|^2\right)
\det(\widetilde{\bm G}+\bm E_G)
}\\
&=-2\ii
\frac{\det(\widetilde{\bm H}+\bm E_H)}
     {\det(\widetilde{\bm G}+\bm E_G)},
\end{aligned}
\label{app:perturbed-effective-ratio}
\end{equation}
where $\bm E_H,\bm E_G=O(e^{-2L_\mu})$.

\paragraph{Denominator bound and uniform error.}
The upper-half-plane condition implies
\[
\int_0^\infty e^{-\ii\kappa_j^*s}e^{\ii\kappa_ms}\,\mathrm ds
=\frac{-\ii}{\kappa_j^*-\kappa_m}.
\]
Hence, for any nonzero vector $\bm w\in\mathbb C^{|\mathcal C_\mu|}$,
\begin{equation}
\begin{aligned}
&\bm w^\dagger(-\ii\widetilde{\bm G})\bm w\\
&\quad=\int_0^\infty\Bigg[\left|\sum_{j\in\mathcal C_\mu}
w_je^{-\widetilde g_{j|\mu}}e^{\ii\kappa_js}\right|^2\\
&\qquad\qquad+\left|\sum_{j\in\mathcal C_\mu}
w_je^{\widetilde g_{j|\mu}}e^{\ii\kappa_js}\right|^2\Bigg]\,\mathrm ds>0,
\end{aligned}
\label{app:effective-gram-positivity}
\end{equation}
where $\dagger$ denotes conjugate transpose. Strict positivity follows from the linear independence of the exponentials for distinct eigenvalues, since $\bm w\ne0$ and $e^{-\widetilde g_{j|\mu}}\ne0$ for every $j$. Hence $\det\widetilde{\bm G}\ne0$.

By Eqs.~\eqref{eq:local-effective-state} and~\eqref{eq:effective-local-g}, $2\widetilde g_{j|\mu}=\mathcal Z_{j|\mu}^{\mathrm{eff}}$. The zero real part and common phase modulo $2\pi$ in Eq.~\eqref{eq:local-sync} therefore imply
\[
\widetilde g_{j|\mu}(P_\mu)
=\frac{\ii\Phi_\mu}{2}+\ii\pi n_{j\mu},
\qquad n_{j\mu}\in\mathbb Z.
\]
The correction in Eq.~\eqref{eq:effective-local-g} is independent of $x,t$. Subtracting the value at $P_\mu$ using Eq.~\eqref{eq:half-exponent} consequently gives
\begin{equation}
\begin{aligned}
\widetilde g_{j|\mu}(x,t)
={}&-\ii\kappa_j(x-X_\mu)-2\ii\kappa_j^2(t-T_\mu)\\
&+\frac{\ii\Phi_\mu}{2}+\ii\pi n_{j\mu}.
\end{aligned}
\label{app:effective-local-coordinates}
\end{equation}
The common phase cancels in $\widetilde G_{jm}$, while the integer branches contribute matching row and column signs, leaving its determinant unchanged. Thus $\det\widetilde{\bm G}$ depends only on the fixed participating spectrum and local coordinates. By continuity and nonvanishing on the compact neighborhood $\mathcal U_\mu$,
\begin{equation}
|\det\widetilde{\bm G}|\ge\eta_\mu>0,
\label{app:denominator-lower-bound}
\end{equation}
where $\eta_\mu$ is independent of $L_\mu$.

Equation~\eqref{app:effective-local-coordinates} also ensures that the entries of $\widetilde{\bm G}$ and $\widetilde{\bm H}$ are uniformly bounded on $\mathcal U_\mu$. Since the determinant is a polynomial in the matrix entries, the perturbations in Eq.~\eqref{app:perturbed-effective-ratio} produce determinant errors of order $O(e^{-2L_\mu})$. Consequently, for sufficiently large $L_\mu$,
\[
|\det(\widetilde{\bm G}+\bm E_G)|
\ge\eta_\mu/2.
\]
Together with Eq.~\eqref{app:denominator-lower-bound}, the determinant perturbation estimates yield
\begin{equation}
q_N(x,t)
=-2\ii\frac{\det\widetilde{\bm H}}
{\det\widetilde{\bm G}}
+O(e^{-2L_\mu})
=q_{\mathcal C_\mu}^{\mathrm{eff}}(x,t)
+O(e^{-2L_\mu}),
\label{app:local-reduction-result}
\end{equation}

\paragraph{Amplitude at the synchronized center.}
At $P_\mu$, Eq.~\eqref{app:effective-local-coordinates} reduces to $\widetilde g_{j|\mu} =\ii\Phi_\mu/2+\ii\pi n_{j\mu}$. After canceling the row and column sign factors in the determinant ratio, we obtain
\begin{equation}
\begin{aligned}
q_{\mathcal C_\mu}^{\mathrm{eff}}(P_\mu)
&=-2\ii
\frac{
\det\begin{pmatrix}
0&e^{\ii\Phi_\mu/2}\bm1^{\mathsf T}\\
e^{\ii\Phi_\mu/2}\bm1
&2\bm K_{\mathcal C_\mu,\mathcal C_\mu}
\end{pmatrix}
}{
\det(2\bm K_{\mathcal C_\mu,\mathcal C_\mu})
}\\
&=\ii e^{\ii\Phi_\mu}
\bm1^{\mathsf T}
\bm K_{\mathcal C_\mu,\mathcal C_\mu}^{-1}\bm1.
\end{aligned}
\label{app:synchronized-field}
\end{equation}

To evaluate this scalar, we exploit the displacement identity of the Cauchy matrix. Let $\bm\Lambda=\operatorname{diag}(\kappa_j)_{j\in\mathcal C_\mu}$ and $\bm K=\bm K_{\mathcal C_\mu,\mathcal C_\mu}$. Since $(\kappa_j^*-\kappa_m)K_{jm}=1$, we have
\begin{equation}
\bm\Lambda^*\bm K-\bm K\bm\Lambda
=\bm1\bm1^{\mathsf T}.
\label{app:cauchy-displacement}
\end{equation}
Note that $\bm1^{\mathsf T}\bm K^{-1}\bm1 =\operatorname{tr}(\bm K^{-1}\bm1\bm1^{\mathsf T})$. Using Eq.~\eqref{app:cauchy-displacement} and the cyclic property of the trace, we obtain
\[
\bm1^{\mathsf T}\bm K^{-1}\bm1
=\operatorname{tr}(\bm\Lambda^*-\bm\Lambda)
=\sum_{j\in\mathcal C_\mu}(\kappa_j^*-\kappa_j)
=-2\ii\sum_{j\in\mathcal C_\mu}\beta_j.
\]
Substitution into Eq.~\eqref{app:synchronized-field}, with $\mathcal A_j=2\beta_j$ from Eq.~\eqref{eq:single-amplitude-velocity}, yields
\[
q_{\mathcal C_\mu}^{\mathrm{eff}}(P_\mu)
=e^{\ii\Phi_\mu}
\sum_{j\in\mathcal C_\mu}\mathcal A_j.
\]
Combining this identity with Eq.~\eqref{app:local-reduction-result} and $\bigl||a|-|b|\bigr|\le|a-b|$ gives
\begin{equation}
|q_N(P_\mu)|
=\sum_{j\in\mathcal C_\mu}\mathcal A_j
+O(e^{-2L_\mu}),
\label{app:local-amplitude-result}
\end{equation}
as stated in Eq.~\eqref{eq:local-amplitude-sum}.

\section{Compatibility system construction and event--spectral correspondence}
\label{app:compatibility-construction}
\label{app:event-spectral-equivalence}

Section~\ref{subsec:compatibility-matrix} gives the compatibility equations and their scattering right-hand sides. Here we detail the blockwise construction of $A_{\Gamma,x}$ and $A_{\Gamma,\phi}$ and make explicit the corresponding ordering of $\mathbf b_\Gamma(\mathbf0)$. We also establish the quotient space correspondence used in the rigidity theorem.

For the position block, use the reference events $\mu_j^0$ chosen in the main text and order the pairs $(j,\mu)$ with
\[
\mu\in\mathcal I_j\setminus\{\mu_j^0\},
\]
and denote the $r$th such pair by $(j_r,\mu_r)$. Then
\[
A_{\Gamma,x}
=
\begin{pmatrix}
C_{11}^{(x)} & \cdots & C_{1K}^{(x)}\\
\vdots & \ddots & \vdots\\
C_{M_x1}^{(x)} & \cdots & C_{M_xK}^{(x)}
\end{pmatrix},
\]
where the $1\times2$ block in the columns associated with event $\nu$ is
\[
C_{r\nu}^{(x)}
=
\begin{cases}
\begin{pmatrix}
1&4\alpha_{j_r}
\end{pmatrix},
&\nu=\mu_r,\\[1mm]
\begin{pmatrix}
-1&-4\alpha_{j_r}
\end{pmatrix},
&\nu=\mu_{j_r}^0,\\[1mm]
\begin{pmatrix}
0&0
\end{pmatrix},
&\text{otherwise}.
\end{cases}
\]
Thus the $r$th row of $A_{\Gamma,x}$ is precisely the coefficient row of the position compatibility equation associated with the pair $(j_r,\mu_r)$.

For the phase block, consider the $\ell$th fundamental cycle
\[
\mathfrak C_\ell
=
(j_1,m_1,\ldots,j_p,m_p,j_1),
\qquad
j_{p+1}=j_1.
\]
Its $1\times2$ block in the columns associated with event $\nu$ is
\[
\begin{aligned}
C_{\ell\nu}^{(\phi)}
={}&\bigl(2(\alpha_{j_a}-\alpha_{j_{a+1}}),\\
&\quad 4\bigl[(\alpha_{j_a}^2-\beta_{j_a}^2)
-(\alpha_{j_{a+1}}^2-\beta_{j_{a+1}}^2)\bigr]\bigr),\\
&\hspace{1em}\nu=m_a\ \text{for some }a\in\{1,\ldots,p\},\\
C_{\ell\nu}^{(\phi)}={}&(0,0),
\quad\nu\notin\{m_1,\ldots,m_p\}.
\end{aligned}
\]
Accordingly,
\[
A_{\Gamma,\phi}
=
\begin{pmatrix}
C_{11}^{(\phi)} & \cdots & C_{1K}^{(\phi)}\\
\vdots & \ddots & \vdots\\
C_{N_{\mathrm{cyc}}1}^{(\phi)}
& \cdots &
C_{N_{\mathrm{cyc}}K}^{(\phi)}
\end{pmatrix}.
\]
The $\ell$th row of $A_{\Gamma,\phi}$ is therefore exactly the coefficient row of the phase-cycle equation \eqref{eq:phase-compatibility-linear}. Stacking the position and phase blocks gives Eq.~\eqref{eq:compatibility-matrix}.

With the same row ordering, the corresponding zero-branch right-hand side is
\begin{equation}
\mathbf b_\Gamma(\mathbf0)
=
\begin{pmatrix}
b_{j_1\mu_1}^{(x)}\\
\vdots\\
b_{j_{M_x}\mu_{M_x}}^{(x)}\\
b_1^{(\phi,0)}\\
\vdots\\
b_{N_{\mathrm{cyc}}}^{(\phi,0)}
\end{pmatrix}.
\label{eq:explicit-zero-branch-rhs}
\end{equation}
Together with Eq.~\eqref{eq:phase-dependent-rhs}, this determines the full right-hand side $\mathbf b_\Gamma(\mathbf n)$.

The construction depends on the chosen reference events $\mu_j^0$ and on the chosen fundamental cycle basis. Different choices correspond to equivalent sets of compatibility equations and therefore leave the solution set and the rank of the compatibility system unchanged.

\subsection{Construction of the correspondence map}

By the local kernel characterization~\eqref{eq:local-rigidity-nullspace} and the construction of $R_\Gamma$ in Eq.~\eqref{eq:spectral-rigidity-system}, $\mathbf u\in\ker R_\Gamma$ if and only if, for every event $\mu$, there exist $\mathbf a_\mu\in\mathbb R^2$ and $\omega_\mu\in\mathbb R$ such that
\begin{equation}
\mathbf u_j=\mathbf a_\mu+\omega_\mu J\mathbf p_j,
\qquad j\in\mathcal C_\mu.
\label{eq:app-event-rigid-motion}
\end{equation}
Each shared soliton has the same vector $\mathbf u_j$ in every event containing it.

For a given $\mathbf u$, the local parameters are unique. Indeed, if two parameter pairs produce the same vectors, then for distinct $j,k\in\mathcal C_\mu$,
\[
(\omega_\mu-\omega_\mu')J(\mathbf p_j-\mathbf p_k)=0.
\]
Since $\mathbf p_j\ne\mathbf p_k$ and $J$ is invertible, $\omega_\mu=\omega_\mu'$, and hence $\mathbf a_\mu=\mathbf a_\mu'$. We therefore recover the event variables by
\begin{equation}
\delta X_\mu=(\mathbf a_\mu)_1,
\qquad
\delta T_\mu=\frac{\omega_\mu}{4},
\qquad
\delta\Phi_\mu=(\mathbf a_\mu)_2.
\label{eq:app-event-parameter-recovery}
\end{equation}

Write
\[
\mathbf u_j=(\xi_j,y_j)^{\mathsf T},
\qquad
\delta s_j=-2\beta_j\xi_j,
\qquad
\delta\vartheta_j=y_j+2\alpha_j\xi_j.
\]
Substituting Eq.~\eqref{eq:app-event-parameter-recovery} into Eq.~\eqref{eq:app-event-rigid-motion}, with $J\mathbf p_j=(\alpha_j,-|\kappa_j|^2)^{\mathsf T}$, gives Eqs.~\eqref{eq:xi-incidence} and \eqref{eq:eta-incidence}. Subtracting the position relation between two events containing the same soliton eliminates $\xi_j$ and gives the homogeneous position compatibility equations. Taking differences of the phase relation within each event eliminates $\delta\Phi_\mu$, and summing around a cycle cancels the soliton phase variations. This gives the homogeneous phase-cycle compatibility equations. Thus
\[
A_\Gamma\delta\mathbf P=0,
\qquad
\delta\mathbf P
=(\delta X_1,\delta T_1,\ldots,\delta X_K,\delta T_K)^{\mathsf T}.
\]

Every $\mathbf u\in\ker R_\Gamma$ therefore determines a unique $\delta\mathbf P\in\ker A_\Gamma$. The recovered parameters depend linearly on $\mathbf u$, because linear combinations of local rigid motions have the corresponding linear combinations of their unique parameters. Hence
\begin{equation}
\Pi:\ker R_\Gamma\longrightarrow\ker A_\Gamma,
\qquad
\mathbf u\longmapsto\delta\mathbf P
\label{eq:kernel-map-Pi}
\end{equation}
is a linear map.

\subsection{Surjectivity of the correspondence map}

Conversely, let $\delta\mathbf P\in\ker A_\Gamma$. The homogeneous position compatibility equations ensure that $\xi_j$ in Eq.~\eqref{eq:xi-incidence} is independent of the choice of event containing soliton $j$. To recover the phase variations, fix $\delta\Phi_{\mu_*}=0$ at one reference event and use Eq.~\eqref{eq:eta-incidence} successively along the soliton--event incidences to determine the remaining phase variations. Connectedness ensures that every soliton and event can be reached. The homogeneous phase-cycle conditions ensure that different paths give the same result, so Eq.~\eqref{eq:eta-incidence} holds on every incidence.

Set
\[
\mathbf u_j=
\begin{pmatrix}
\xi_j\\
\delta\vartheta_j-2\alpha_j\xi_j
\end{pmatrix}.
\]
Equations~\eqref{eq:xi-incidence} and \eqref{eq:eta-incidence} give
\[
\mathbf u_j=
\begin{pmatrix}
\delta X_\mu\\
\delta\Phi_\mu
\end{pmatrix}
+4\delta T_\mu J\mathbf p_j,
\qquad j\in\mathcal C_\mu.
\]
Thus $\mathbf u\in\ker R_\Gamma$ and $\Pi(\mathbf u)=\delta\mathbf P$. Since $\delta\mathbf P$ was arbitrary, $\Pi$ is surjective:
\begin{equation}
\Pi(\ker R_\Gamma)=\ker A_\Gamma.
\label{eq:Pi-surjective}
\end{equation}

\subsection{Correspondence of trivial motions}

Let $\mathcal T_A$ and $\mathcal T_R$ be the trivial motion spaces defined in Theorem~\ref{thm:rigidity-balance}. For a global spectral motion $\mathbf u_j=\mathbf a+\omega J\mathbf p_j$, with $\mathbf a=(a_1,a_2)^{\mathsf T}$, Eq.~\eqref{eq:app-event-parameter-recovery} gives
\[
\delta X_\mu=a_1,
\qquad
\delta T_\mu=\omega/4,
\qquad
\delta\Phi_\mu=a_2
\qquad\text{for all }\mu.
\]
Hence every global spectral motion produces a uniform event translation. Conversely, any uniform event translation $\delta X_\mu=a$, $\delta T_\mu=b$ is obtained by choosing $\mathbf a=(a,0)^{\mathsf T}$ and $\omega=4b$. Therefore,
\begin{equation}
\Pi(\mathcal T_R)=\mathcal T_A.
\label{eq:Pi-trivial-image}
\end{equation}

To determine $\ker\Pi$, suppose $\Pi(\mathbf u)=0$. Then $\delta X_\mu=\delta T_\mu=0$ for every event, and Eq.~\eqref{eq:app-event-rigid-motion} reduces to
\[
\mathbf u_j=(0,\delta\Phi_\mu)^{\mathsf T},
\qquad j\in\mathcal C_\mu.
\]
Events sharing a soliton must have equal $\delta\Phi_\mu$. Connectedness therefore makes all event phase variations equal. Conversely, any common vertical translation gives zero event displacement. Thus, with $\mathbf e_2=(0,1)^{\mathsf T}$,
\begin{equation}
\ker\Pi
=
\left\{
\mathbf u:\mathbf u_j=c\mathbf e_2
\text{ for all }j,\quad c\in\mathbb R
\right\}
\subseteq\mathcal T_R.
\label{eq:Pi-kernel}
\end{equation}

If $\Pi(\mathbf u)\in\mathcal T_A$, then Eq.~\eqref{eq:Pi-trivial-image} provides $\mathbf v\in\mathcal T_R$ with $\Pi(\mathbf v)=\Pi(\mathbf u)$. Consequently, $\mathbf u-\mathbf v\in\ker\Pi\subseteq\mathcal T_R$, so $\mathbf u\in\mathcal T_R$. Together with Eq.~\eqref{eq:Pi-trivial-image}, this proves
\begin{equation}
\Pi^{-1}(\mathcal T_A)=\mathcal T_R.
\label{eq:Pi-preimage-trivial}
\end{equation}

\subsection{Quotient space correspondence}

Since $\mathcal T_R$ and $\mathcal T_A$ are linear subspaces of $\ker R_\Gamma$ and $\ker A_\Gamma$, respectively, the corresponding quotient spaces are well defined.

Consider the induced map
\begin{equation}
\overline{\Pi}:
\ker R_\Gamma/\mathcal T_R
\longrightarrow
\ker A_\Gamma/\mathcal T_A,
\qquad
[\mathbf u]\longmapsto[\Pi(\mathbf u)].
\label{eq:quotient-deformation-correspondence}
\end{equation}
If $\mathbf u-\mathbf v\in\mathcal T_R$, then Eq.~\eqref{eq:Pi-trivial-image} implies $\Pi(\mathbf u)-\Pi(\mathbf v)\in\mathcal T_A$. Thus $\overline{\Pi}$ is well defined. It inherits linearity and surjectivity from $\Pi$.

To prove injectivity, suppose $\overline{\Pi}([\mathbf u])=[0]$. Then $\Pi(\mathbf u)\in\mathcal T_A$, and Eq.~\eqref{eq:Pi-preimage-trivial} gives $\mathbf u\in\mathcal T_R$, hence $[\mathbf u]=[0]$. Therefore $\overline{\Pi}$ is a linear isomorphism:
\begin{equation}
\ker R_\Gamma/\mathcal T_R
\cong
\ker A_\Gamma/\mathcal T_A.
\label{eq:quotient-deformation-correspondence-final}
\end{equation}
Taking dimensions gives
\[
\dim\ker R_\Gamma-3
=
\dim\ker A_\Gamma-2
=
D_{\mathrm{int}},
\]
which establishes the internal deformation correspondence in Theorem~\ref{thm:rigidity-balance}.

\section{Explicit rigidity counts for representative complete organizations}
\label{app:explicit-rigidity-counts}

\subsection{Fully pairwise organizations}
\label{app:complete-pairwise-proof}

For the fully pairwise organization, every unordered pair of spectral points contributes the constraint
\begin{equation}
(\mathbf p_i-\mathbf p_j)^{\mathsf T}
(\mathbf u_i-\mathbf u_j)=0,
\qquad
1\leq i<j\leq N.
\label{eq:pairwise-appendix-constraint}
\end{equation}

If the spectral points are not all collinear, Corollary~\ref{cor:complete-rigidity} gives
\[
D_{\mathrm{int}}=0,
\qquad
\dim\ker R_\Gamma=3.
\]

Now suppose that all spectral points are collinear. Let $\mathbf t$ be a unit tangent vector to their common line and $\widehat{\mathbf n}$ a unit normal vector, and write
\[
\mathbf u_j
=
a_j\mathbf t+b_j\widehat{\mathbf n}.
\]
Since every difference $\mathbf p_i-\mathbf p_j$ is parallel to $\mathbf t$, Eq.~\eqref{eq:pairwise-appendix-constraint} implies
\[
a_i=a_j
\qquad
\text{for all }i,j,
\]
while the normal components $b_1,\ldots,b_N$ remain unrestricted. Hence
\[
\dim\ker R_\Gamma=N+1,
\qquad
D_{\mathrm{int}}=N-2.
\]

For the fully pairwise organization, each event has two incidences, so $N_{\mathrm{inc}}=2K$. Since $K=\binom{N}{2}$, Eq.~\eqref{eq:constraint-surplus} gives
\[
\Delta_\Gamma
=
K-2N+3
=
\frac{(N-2)(N-3)}{2}.
\]
Therefore, using $\rho_\Gamma=\Delta_\Gamma+D_{\mathrm{int}}$,
\[
\rho_\Gamma
=
\begin{cases}
\displaystyle
\frac{(N-2)(N-3)}{2},
& \begin{gathered}\text{if the spectral points}\\\text{are not all collinear}\end{gathered},\\[2mm]
\displaystyle
\frac{(N-1)(N-2)}{2},
& \begin{gathered}\text{if the spectral points}\\\text{are all collinear}\end{gathered}.
\end{cases}
\]
This proves Corollary~\ref{cor:pairwise-rigidity}.

\subsection{One collective event with pairwise remainder}
\label{app:mixed-collective-pairwise-proof}

Consider a complete organization with one $m$-soliton collective event and all remaining interactions pairwise, where $3\leq m\leq N$. Let $\mathcal C_*$ denote the set of solitons participating in the collective event. The collective-event constraint implies all pairwise rigidity constraints among the solitons in $\mathcal C_*$. Since all remaining interactions are pairwise, every $\mathbf u\in\ker R_\Gamma$ satisfies the fully pairwise rigidity constraints.

If the auxiliary spectral points are not all collinear, Corollary~\ref{cor:complete-rigidity} gives
\[
D_{\mathrm{int}}=0.
\]

Now suppose that all auxiliary spectral points are collinear. Let $\mathbf t$ be a unit tangent vector to their common line and set $\widehat{\mathbf n}=J\mathbf t$. Write
\[
\mathbf u_j
=
a_j\mathbf t+b_j\widehat{\mathbf n}.
\]
Since the collective-event constraint supplies the pairwise rigidity constraints within $\mathcal C_*$, the full set of pairwise rigidity constraints gives
\[
a_1=\cdots=a_N=:a_t,
\]
while imposing no restriction on the normal components.

For $j\in\mathcal C_*$, write $\mathbf p_j=\mathbf p_0+\lambda_j\mathbf t$. The collective-event condition gives
\[
\mathbf u_j
=
\mathbf a_*+\omega_*J\mathbf p_j
=
\mathbf a_*+\omega_*J\mathbf p_0
+\omega_*\lambda_j\widehat{\mathbf n}.
\]
Since the last term is purely normal and all tangential components equal $a_t$, write
\[
\mathbf a_*+\omega_*J\mathbf p_0
=
a_t\mathbf t+c\widehat{\mathbf n}.
\]
Hence
\[
\mathbf u_j
=
a_t\mathbf t
+
(c+\omega_*\lambda_j)\widehat{\mathbf n},
\qquad
j\in\mathcal C_*.
\]
The quantities $\lambda_j$ are fixed by the spectral configuration, so the normal components of the $m$ core points depend only on $c$ and $\omega_*$. For $r\notin\mathcal C_*$, the pairwise constraints leave the normal component $b_r$ free. Thus the free parameters are $a_t$, $c$, $\omega_*$, and $(b_r)_{r\notin\mathcal C_*}$, giving
\[
\begin{aligned}
\dim\ker R_\Gamma&=1+2+(N-m)=N-m+3,\\
D_{\mathrm{int}}&=N-m.
\end{aligned}
\]

For this organization,
\[
K
=
1+\binom{N}{2}-\binom{m}{2},
\qquad
N_{\mathrm{inc}}
=
m+2(K-1).
\]
Equation~\eqref{eq:constraint-surplus} therefore gives
\[
\Delta_\Gamma
=
\frac{(N-m)(N+m-5)}{2}.
\]
Using $\rho_\Gamma=\Delta_\Gamma+D_{\mathrm{int}}$, we obtain
\[
\rho_\Gamma
=
\begin{cases}
\displaystyle
\frac{(N-m)(N+m-5)}{2},
& \begin{gathered}\text{if the spectral points}\\\text{are not all collinear}\end{gathered},\\[2mm]
\displaystyle
\frac{(N-m)(N+m-3)}{2},
& \begin{gathered}\text{if the spectral points}\\\text{are all collinear}\end{gathered}.
\end{cases}
\]

This proves Corollary~\ref{cor:mixed-collective-pairwise}.

\section{Matrix reduction of the
\texorpdfstring{$3$--$2$--$2$--$2$}{3-2-2-2} pattern}
\label{app:3222-reduction}

For $\Gamma_{\mathrm m}=\{123,14,24,34\}$, fix $P_{123}=(0,0)$ and let $\widehat A_{\mathrm m,x}$ and $\widehat A_{\mathrm m,\phi}$ be the gauge-fixed blocks obtained from the general construction in Appendix~\ref{app:compatibility-construction}. There are five position rows and two independent phase-cycle rows. For distinct velocities, the minor of $\widehat A_{\mathrm m,x}$ obtained by deleting the $T_{14}$ column is
\[
-16(\alpha_2-\alpha_4)(\alpha_3-\alpha_4)\ne0,
\]
so $\operatorname{rank}\widehat A_{\mathrm m,x}=5$.

Let $\mathbf P_{\mathrm m,0}$ be the unique solution of $\widehat A_{\mathrm m,x}\mathbf P=\mathbf b_{\mathrm m,x}$ with $T_{14}=0$, and let $\mathbf v$ span $\ker\widehat A_{\mathrm m,x}$ with $v_2=1$. Then $\mathbf P_{\mathrm m}=\mathbf P_{\mathrm m,0}+\mathbf vT$, where $T=T_{14}$.

Choose the oriented cycles $\mathfrak C_{124}=1-123-2-24-4-14-1$ and $\mathfrak C_{134}=1-123-3-34-4-14-1$, and write
\[
\widehat A_{\mathrm m,\phi}\mathbf P_{\mathrm m,0}
=
\begin{pmatrix}
Q_{124}\\
Q_{134}
\end{pmatrix},
\qquad
\mathbf b_{\mathrm m,\phi}(\mathbf n)
=
\begin{pmatrix}
\Psi_{124}\\
\Psi_{134}
\end{pmatrix}.
\]
Restricting the phase block to the affine position family gives
\[
\begin{aligned}
G_{\mathrm m}T&=\mathbf g_{\mathrm m},\\
G_{\mathrm m}&=\widehat A_{\mathrm m,\phi}\mathbf v,\\
\mathbf g_{\mathrm m}&=\mathbf b_{\mathrm m,\phi}(\mathbf n)
-\widehat A_{\mathrm m,\phi}\mathbf P_{\mathrm m,0}.
\end{aligned}
\]
Direct evaluation of the two phase rows yields
\[
G_{\mathrm m}
=
4
\begin{pmatrix}
d_{124}\\
d_{134}
\end{pmatrix},
\]
and hence Eq.~\eqref{eq:3222-main-reduced-system}.

\section{Parameters and admissibility of the mixed flexible family}
\label{app:mixed-numerical-construction}

\subsection{Spectral parameters}

For the mixed family $\Gamma_{\mathrm m}=\{123,14,24,34\}$ used in the full-field and Fourier examples, set $P_{123}=(0,0)$ and fix
\begin{equation}
\begin{aligned}
&(\epsilon_4^{(123)},\epsilon_2^{(14)},\epsilon_3^{(14)},
\epsilon_1^{(24)},\epsilon_3^{(24)},\epsilon_1^{(34)},\epsilon_2^{(34)})\\
&\qquad=(-1,+1,+1,-1,+1,-1,-1),
\end{aligned}
\label{eq:mixed-numerical-sectors}
\end{equation}
with integer branch $(n_{124},n_{134})=(-1,-2)$.

We seek a unit-circle spectrum of the form
\begin{equation}
\kappa_j^{(\mathrm f)}=e^{\ii\theta_j},\qquad
\theta_1=\frac{\pi}{4},\qquad
\theta_4=\frac{3\pi}{4}.
\label{eq:flexible-circle-family}
\end{equation}
Let $u:=\theta_2$ and $v:=\theta_3$, with $\pi/4<u<v<3\pi/4$.

For the prescribed ordering, the principal logarithm in Eq.~\eqref{eq:scattering-log-principal} gives
\[
\operatorname{Im}\Omega_{jm}
=
\begin{cases}
\theta_m-\pi,&j<m,\\
\theta_m,&j>m.
\end{cases}
\]
The corresponding real scattering shifts are
\begin{equation}
r_{jm}:=\operatorname{Re}\Omega_{jm}
=\log\frac{\sin[(\theta_m-\theta_j)/2]}
{\sin[(\theta_j+\theta_m)/2]},
\qquad j<m.
\label{eq:flexible-real-shifts}
\end{equation}

For the unit-circle spectrum, the auxiliary spectral points $\mathbf p_j=(1,\alpha_j)$ are collinear, so $d_{124}=d_{134}=0$. A direct evaluation of Eq.~\eqref{eq:phase-compatibility-rhs} using the sectors in Eq.~\eqref{eq:mixed-numerical-sectors} and the cycle orientations of Appendix~\ref{app:3222-reduction} gives $b_{124}^{(\phi,0)}=\pi$ and $b_{134}^{(\phi,0)}=2\pi$. With $\Psi_{abc}=b_{abc}^{(\phi,0)}+2\pi n_{abc}$ and $(n_{124},n_{134})=(-1,-2)$, the compatibility conditions \eqref{eq:3222-degenerate-selection} reduce to
\begin{equation}
Q_{124}=-\pi,\qquad Q_{134}=-2\pi.
\label{eq:flexible-resonance-construction}
\end{equation}

Substituting the position-compatible event centers into these conditions yields $F_1(u,v)=F_2(u,v)=0$, where $F_1:=Q_{124}+\pi$ and $F_2:=Q_{134}+2\pi$ are given by
\begin{equation}
\begin{aligned}
F_1(u,v)&=\pi-\log2+r_{12}+r_{13}+r_{24}\\
&\quad+\cot u\,(r_{12}-r_{23}-r_{24}),\\
F_2(u,v)&=2\pi-\log2+r_{12}+r_{13}+2r_{24}+r_{34}\\
&\quad+\cot v\,(r_{13}+r_{23}-r_{34}).
\end{aligned}
\label{eq:flexible-explicit-closure}
\end{equation}
The selected root has numerical values $\theta_2\approx1.1697116482$ and $\theta_3\approx1.2716864478$.

The exact spectrum is specified by the unique common zero of $F_1$ and $F_2$ in
\[
\begin{aligned}
\mathcal R&=(u_0,v_0)+[-h,h]^2,\\
(u_0,v_0)&=(1.16971165,1.27168645),\quad h=10^{-8}.
\end{aligned}
\]
Throughout $\mathcal R$, the explicit expressions give
\begin{equation}
\begin{aligned}
\partial_uF_1&>4.6,\qquad |\partial_vF_1|<2.4,\\
|\partial_uF_2|&<2.2,\qquad \partial_vF_2>8.5.
\end{aligned}
\label{eq:flexible-elementary-derivative-bounds}
\end{equation}
The center residuals satisfy
\begin{equation}
|F_1(u_0,v_0)|<4\times10^{-9},
\qquad
|F_2(u_0,v_0)|<2\times10^{-8}.
\label{eq:flexible-center-residuals}
\end{equation}
For each $v\in[v_0-h,v_0+h]$, the mean value theorem and the bounds in Eqs.~\eqref{eq:flexible-elementary-derivative-bounds} and~\eqref{eq:flexible-center-residuals} give
\[
F_1(u_0-h,v)<0<F_1(u_0+h,v).
\]
By continuity and strict monotonicity in $u$, the equation $F_1(u,v)=0$ has a unique solution $u=u(v)$ for each $v\in[v_0-h,v_0+h]$. Since $\partial_uF_1>0$, the implicit function theorem ensures that $u(v)$ is continuously differentiable.

Define $H(v)=F_2(u(v),v)$. Implicit differentiation gives
\[
H'(v)
=\partial_vF_2
-\frac{\partial_uF_2\,\partial_vF_1}
{\partial_uF_1},
\]
where the derivatives are evaluated at $(u(v),v)$. The derivative bounds in Eq.~\eqref{eq:flexible-elementary-derivative-bounds} give $H'(v)>7$. Moreover, $F_1(u(v_0),v_0)=0$ and the mean value theorem yield
\[
\begin{aligned}
|H(v_0)|
&\le |F_2(u_0,v_0)|
+\sup_{\mathcal R}|\partial_uF_2|\,
  |u(v_0)-u_0|\\
&\le |F_2(u_0,v_0)|
+\frac{\sup_{\mathcal R}|\partial_uF_2|}
       {\inf_{\mathcal R}\partial_uF_1}
 |F_1(u_0,v_0)|\\
&<2.2\times10^{-8}<7h.
\end{aligned}
\]
Consequently,
\[
H(v_0-h)<H(v_0)-7h<0
<H(v_0)+7h<H(v_0+h).
\]
Since $H$ is strictly increasing, it has exactly one zero, yielding a unique common zero of $F_1$ and $F_2$ in $\mathcal R$.

At this root, the phase-closure conditions are satisfied. Since $d_{124}=d_{134}=0$ for the unit-circle spectrum, the reduced phase system \eqref{eq:3222-main-reduced-system} imposes no restriction on $T_{14}$, establishing algebraic flexibility at fixed spectrum. The admissible deformation interval is determined by the sector, isolation, and event-separation conditions considered below.

\subsection{Reconstruction and admissibility}

The common position norming data and phases follow from Eqs.~\eqref{eq:required-position-norming}--\eqref{sj} and~\eqref{eq:phase-incidence}. At both endpoints $T_{14}=-30$ and $-20$, the sector conditions hold and the event times satisfy $T_{34}<T_{24}<T_{14}<T_{123}=0$, with a minimum separation greater than $2.9980$.

Bounding the scattering shifts gives $2\epsilon_r^{(\mu)}\operatorname{Re}g_r(P_\mu)-C_r>15.5765$ at both endpoints for every $\mu$ and $r\in\mathcal R_\mu$, where $C_r=\sum_{\ell\ne r}|\operatorname{Re}\Omega_{r\ell}|$. For the spectrum in Eq.~\eqref{eq:flexible-circle-family}, with $\theta_2$ and $\theta_3$ determined above, the signed remote coordinates and successive event-time gaps are affine in $T_{14}$, while $C_r$ is constant. These bounds therefore hold throughout $T_{14}\in[-30,-20]$ and give
\[
\max_{\mu,\;r\in\mathcal R_\mu}
\operatorname{sech}\left|
\operatorname{Re}\mathcal Z_{r|\mu}^{\mathrm{loc}}(P_\mu)
\right|
<3.44\times10^{-7}.
\]
Thus the family is physically admissible throughout this interval.

\section{Cluster identities, phase rates, and Fourier comparison}
\label{app:fixed_time_clusters}

\subsection{Centered synchronized cores}
\label{app:fourier-core-symmetry}

We first show that, up to an overall phase, the centered profile of a synchronized effective cluster is determined by its participating eigenvalues. Let $q_{\mathcal C}^{\mathrm{sync}}(\xi)$ be the zero-time solution with sub-spectrum $(\kappa_j)_{j\in\mathcal C}$ and $c_j=0$ for all $j\in\mathcal C$. At the event center $(X_\mu,T_\mu)$, the synchronization conditions~\eqref{eq:local-sync} imply $\mathcal Z_{j|\mu}^{\mathrm{eff}}(X_\mu,T_\mu) =\ii\Phi_\mu\pmod{2\pi\ii}$ for all $j\in\mathcal C_\mu$. Setting $x=X_\mu+\xi$ at $t=T_\mu$ gives
\begin{equation}
\begin{aligned}
&\mathcal Z_{j|\mu}^{\mathrm{eff}}(X_\mu+\xi,T_\mu)\\
&\quad=2\beta_j\xi+\ii(\Phi_\mu-2\alpha_j\xi)
\pmod{2\pi\ii},\quad j\in\mathcal C_\mu.
\end{aligned}
\label{eq:fourier-centered-effective-state}
\end{equation}
The determinant representation then gives
\begin{equation}
q_{\mathcal C_\mu}^{\mathrm{eff}}(X_\mu+\xi,T_\mu)
=e^{\ii\Phi_\mu}q_{\mathcal C_\mu}^{\mathrm{sync}}(\xi).
\label{eq:fourier-centered-cluster-identity}
\end{equation}

For the standard zero-time synchronized profile, $g_j(\xi)=-\ii\kappa_j\xi$. Let $f_j=e^{g_j}$ and $h_j=e^{-g_j}$, and let $\bm G$ be the denominator matrix restricted to the cluster. Its entries give $\bm G^\dagger=-\bm G$, while $\xi\mapsto-\xi$ exchanges $f$ and $h$ and leaves $\bm G$ unchanged. The Schur complement representation is
\begin{equation}
q_{\mathcal C}^{\mathrm{sync}}(\xi)
=2\ii f^{\mathsf T}\bm G^{-1}\overline h.
\end{equation}
Taking its complex conjugate and transposing the resulting scalar, using $\overline{\bm G^{-1}}^{\mathsf T}=-\bm G^{-1}$, gives $\overline{q_{\mathcal C}^{\mathrm{sync}}(\xi)} =2\ii h^{\mathsf T}\bm G^{-1}\overline f =q_{\mathcal C}^{\mathrm{sync}}(-\xi)$. A change of variable in the Fourier integral proves
\begin{equation}
\begin{aligned}
q_{\mathcal C}^{\mathrm{sync}}(-\xi)
&=\overline{q_{\mathcal C}^{\mathrm{sync}}(\xi)},\\
\mathcal F_{\mathcal C}(k)&:=\int_{\mathbb R}
q_{\mathcal C}^{\mathrm{sync}}(\xi)e^{-\ii k\xi}\,\dd\xi\in\mathbb R.
\end{aligned}
\label{eq:fourier-core-symmetry}
\end{equation}
Real-valuedness does not imply nonnegativity or evenness in $k$.

\subsection{Rates of the mixed-family distance and phase}
\label{app:fourier-phase-rates}

We derive the rates in Eq.~\eqref{eq:fourier-exact-rates} from the position reconstruction relations \eqref{eq:required-position-norming} and \eqref{sj}, together with the phase synchronization condition \eqref{eq:phase-incidence}. The shifts $\Omega_{jm}$ remain constant along the family because the spectrum and sector assignment are fixed.

At $P_{123}=(0,0)$, the sector sign is $\epsilon_4^{(123)}=-1$, so Eqs.~\eqref{eq:required-position-norming} and \eqref{sj} give $s_1=\operatorname{Re}\Omega_{14}$. At the event $P_{14}=(X_{14},\tau)$, the remote signs are $\epsilon_2^{(14)}=\epsilon_3^{(14)}=+1$. The same formula gives
\[
s_1=-2\beta_1(X_{14}-V_1\tau)
-\operatorname{Re}(\Omega_{12}+\Omega_{13}).
\]
Equating the two expressions for the same $s_1$ yields
\begin{equation}
\begin{aligned}
X_{14}(\tau)&=V_1\tau
-\frac{\operatorname{Re}(\Omega_{12}+\Omega_{13}+\Omega_{14})}{2\beta_1},\\
X_{14}'&=V_1=-4\alpha_1.
\end{aligned}
\label{app:eq:fourier-affine-X14}
\end{equation}

Next, using Eqs.~\eqref{eq:required-position-norming} and \eqref{sj} for soliton~$4$ at $P_{14}$ gives
\[
s_4(\tau)
=
-2\beta_4(X_{14}(\tau)-V_4\tau)
-\operatorname{Re}(\Omega_{42}+\Omega_{43}).
\]
Using Eq.~\eqref{eq:fourier-effective-norming} and $\operatorname{Re}c_4=s_4$, we obtain
\[
\begin{aligned}
\operatorname{Re}c_4^{\mathrm{eff}}(\tau)
&=s_4(\tau)+\operatorname{Re}H_4\\
&=-2\beta_4(X_{14}(\tau)-V_4\tau)
+\operatorname{Re}\Omega_{41}.
\end{aligned}
\]
The center formula in Eq.~\eqref{eq:fourier-distance-phase} therefore becomes
\begin{equation}
\begin{aligned}
D(\tau)&=X_{14}(\tau)-V_4\tau
-\frac{\operatorname{Re}\Omega_{41}}{2\beta_4},\\
D'&=V_1-V_4=4(\alpha_4-\alpha_1).
\end{aligned}
\label{app:eq:fourier-distance-intercept}
\end{equation}
This proves the affine dependence of the cluster separation on $\tau$ without introducing an additional intercept parameter. For the spectrum~\eqref{eq:flexible-circle-family},
\[
D(\tau)\approx -4\sqrt{2}\,\tau+2.425734,
\]
which is positive throughout $\tau\in[-30,-20]$.

On the fixed integer branch, synchronization of the $123$ core with $\Phi_{123}=0$ gives $\vartheta_1=\operatorname{Im}\Omega_{14}$. At the event $14$, Eq.~\eqref{eq:phase-incidence} for soliton~$1$ reads
\[
\begin{aligned}
\Phi_{14}={}&\vartheta_1-2\alpha_1X_{14}
-4(\alpha_1^2-\beta_1^2)\tau\\
&+\operatorname{Im}(\Omega_{12}+\Omega_{13})\pmod{2\pi}.
\end{aligned}
\]
Differentiation, together with $X_{14}'=V_1=-4\alpha_1$, gives
\begin{equation}
\Phi_{14}'=-2\alpha_1X_{14}'
-4(\alpha_1^2-\beta_1^2)=4|\kappa_1|^2.
\label{app:eq:fourier-Phi14-rate}
\end{equation}
The same phase-incidence equation for soliton~$4$ gives
\[
\begin{aligned}
\vartheta_4={}&\Phi_{14}+2\alpha_4X_{14}
+4(\alpha_4^2-\beta_4^2)\tau\\
&-\operatorname{Im}(\Omega_{42}+\Omega_{43})\pmod{2\pi}.
\end{aligned}
\]

Since $H_4$ is independent of $\tau$, Eq.~\eqref{eq:fourier-effective-norming} gives
\begin{equation}
\frac{\dd}{\dd\tau}\operatorname{Im}c_4^{\mathrm{eff}}
=\vartheta_4'
=4\bigl[|\kappa_1|^2-2\alpha_1\alpha_4
+\alpha_4^2-\beta_4^2\bigr].
\label{app:eq:fourier-norming-phase-rate}
\end{equation}
Finally, differentiating the centered phase in Eq.~\eqref{eq:fourier-distance-phase} and substituting the separation rate gives
\[
\phi'=\vartheta_4'-2\alpha_4D'
=4(|\kappa_1|^2-|\kappa_4|^2),
\]
which proves Eq.~\eqref{eq:fourier-exact-rates}.

\subsection{Numerical comparison of the two-cluster prediction}
\label{app:fourier-numerical-verification}

For \(\tau=-30,-25,-20\), we compare the Fourier power spectra of the full four-soliton fields with the two-cluster prediction, using \(D\) and \(\phi\) reconstructed from the norming data. Using a zero-padded FFT of length \(2^{18}\) on \(x\in[-60,240)\) with \(\Delta x=0.02\), we obtain \(\max_{|k|\leq20}|E|<10^{-10}\) and \(\max_{k\in I_k}|\mathcal I-\cos\Psi|<10^{-11}\) on the sampled frequency grid, where \(I_k=[1.6,2.4]\). Halving \(\Delta x\) or extending the window to \([-80,260)\) changes the complex Fourier transforms by less than \(10^{-11}\) over the sampled frequency range.

\bibliography{ref}

\end{document}